\RequirePackage{fixltx2e}
\documentclass[preprintnumbers,prb,superscriptaddress,preprint]{revtex4-1}
\usepackage{amsmath}
\usepackage{graphicx}
\usepackage{dcolumn}
\usepackage{latexsym,bm}
\usepackage{hyperref}
\usepackage{amsfonts}
\usepackage{tabularx}
\usepackage{cases}
\usepackage{amssymb}
\usepackage{enumerate}
\usepackage{mathtools}
\usepackage{etoolbox} 
\usepackage{lipsum} 
\usepackage[capitalize]{cleveref}
\usepackage{xcolor}
\usepackage[normalem]{ulem}

\begin{document}

\title{Modulation of Nearly Free Electron States in Hydroxyl-Functionalized MXenes: A First-Principles Study}

\author{Jiaqi Zhou}
\affiliation{Department of Physics, Shanghai Normal University, Shanghai 200234, China}

\author{Mohammad Khazaei}
\affiliation{Department of Physics, Yokohama National University, Yokohama 240-8501, Japan}

\author{Ahmad Ranjbar}
\affiliation{Dynamics of Condensed Matter and Center for Sustainable Systems Design, Chair of Theoretical Chemistry, University of Paderborn, Warburger Str. 100, D-33098 Paderborn, Germany}

\author{Vei Wang}
\affiliation{Department of Applied Physics, Xi’an University of Technology, Xi’an 710054, China}

\author{Thomas D. K\"{u}hne}
\affiliation{Dynamics of Condensed Matter and Center for Sustainable Systems Design, Chair of Theoretical Chemistry, University of Paderborn, Warburger Str. 100, D-33098 Paderborn, Germany}

\author{Kaoru Ohno}
\affiliation{Department of Physics, Yokohama National University, Yokohama 240-8501, Japan}

\author{Yoshiyuki Kawazoe}
\affiliation{New Industry Creation Hatchery Center, Tohoku University, Sendai, 980-8579, Japan}
 \affiliation{School of Physics, Institute of Science and Center of Excellence in Advanced Functional Materials, Suranaree University of Technology, Nakhon Ratchasima 30000, Thailand}

\author{Yunye Liang}
\email{E-mail: liangyunye@shnu.edu.cn}
\affiliation{Department of Physics, Shanghai Normal University, Shanghai 200234, China}

\date{\today}

\widetext

\begin{abstract}

The two-dimensional transition metal carbides and nitrides (named as MXenes) and their functionalized ones exhibit various physical and chemical characteristics. For example, it has been reported that the nearly free electron (NFE) states can be energetically found near the Fermi levels in hydroxyl functionalized MXenes. Most of these OH-terminated MXene are metallic, but some of them, e.g. Sc$_2$C(OH)$_2$, are semiconductors with NFE states conduction bands. In a variety of low dimensional materials, such as graphene, BN nanotubes and fullerenes, NFE states have been theoretically predicted and/or observed experimentally. In these systems, NFE states play less important roles in chemical reactions or electronic device applications because they appear at energies several electron-volts above the Fermi level. Here, based on the density functional theory (DFT) calculations and the image-potential well model, we show that the wave functions of these NFE states are spatially extensive outside the surface. We propose that the energy gap width is affected by the interlayer distance because of the significant overlap and the hybridization between the wave functions of NFE states from the neighboring layers. We also demonstrated that the energetics of the NFE bands can be engineered by the external electric fields. This results in semiconducting to metallic transition in Sc$_2$C(OH)$_2$. The band-gap manipulation makes Sc$_2$C(OH)$_2$ an excellent candidate for electronic switch applications. Finally, by performing a set of electron transport calculations, \textit{I-V} characteristics of Sc$_2$C(OH)$_2$ device is investigated at various gate voltages.
It is illustrated the NFE states in Sc$_2$C(OH)$_2$ contribute in the transport properties significantly.
\end{abstract}
\bigskip


\maketitle
\section{Introduction}

The successful synthesis of graphene has triggered a significant research interest in synthesis and applications of two-dimensional (2D) materials due to their large surface area and unique physical and chemical properties.~\cite{K. S. Novoselov} Since then, various 2D materials, such as silicene, transition-metal dichalcogenides (TMDCs), and black phosphorus have been synthesized and studied.~\cite{B. Aufray, P. Vogt, B. Radisavljevic,K. F. Mark, J. N. Coleman, Q. H. Wang,L. Rapoport,Li Likai} These materials are regarded as prospective candidates for the next-generation nano-electronic devices. More recently, a novel family of 2D transition metal carbides and nitrides, named as MXenes, have been exfoliated successfully from the MAX phases and attracted great attentions.~\cite{M. Naguib, M. Naguib2, M. Naguib3, M. Khazaei, M. Khazaei2, E.Balci2018, M.Khazaei2017, M.Khazaei2019, B.Anasori2017_1, B.M.Jun2018, J.Pang2019, A.L.Ivanovskii2013, N.K.Chaudhari2017, J.Zhu2017,H.Wang2018,X.Li2018,X.Zhang2018,Y.Zhang2018,K.Hantanasirisakul2018,H.Lin2018,C.Zhan2019} The MAX materials have a general chemical formula of M$_{n+1}$AX$_n$, where M stands for the transition metal, A represents IIIA or IVA elements (such as Al, Ga, Si or Ge), X is C or N, and $n$ = 1$-$3.~\cite{M.Khazaei2014} Because the bonds between M and A atoms are weaker than those between M and X atoms, M$_{n+1}$X$_n$  layers can be readily obtained by etching of A atoms using acid treatment.~\cite{M. Naguib,M.Khazaei2018}

MXenes have some unique properties. Without surface functionalization, most of them are metallic. But, upon proper surface termination with -OH, -F or -O groups, some of the functionalized MXenes become semiconductors.~\cite{M. Khazaei} Based on the first-principles calculations, the nearly free electron (NFE) states in MXene films has been reported, recently.~\cite{NFE} The NFE states, stemming from the image-potential (IP) states, have been predicted and observed in different low-dimensional materials.~\cite{J. Zhao,J. Zhao2,M. Feng,M. Feng2,S. Bose} For example, the hydrogen-like IP states have been measured in fullerenes experimentally and the hybridization of these IP orbitals forms NFE bands on fullerites.~\cite{M. Feng,M. Feng2} In OH-terminated MXene monolayers, their NFE bands are very close to the Fermi levels.~\cite{NFE} This behavior is distinct from other monolayers, such as graphene, graphane, MoS$_2$ and other MXenes terminated with -O or -F groups whose NFE states are several electron-volts away from the Fermi levels.~\cite{NFE,J. Zhao,V. M. Silkin,N. T. Cuong} For this reason, the NFE bands in OH-terminated MXenes can be used as the channel for the electron transport.~\cite{NFE,J. Zhao}

In this paper, the properties of the NFE states and their modulation are investigated. It has been proposed that NFE bands can be modulated by applying electric fields or charge doping.~\cite{K. H. Khoo,M. Ishigami,J. Zhao charging,E. R. Margine, E. Balci} Here, we show that the energetics of the NFE states can be engineered by controlling the interlayer distances between monolayers. This results in band gap tuning. The band gaps are also sensitive to the external electric fields. The wave functions of the NFE states are extensive and  spatially parallel to the surface. These imply that they can be ideal electron transport channels without being scattered from surfaces.~\cite{NFE} The modulation indicates that these materials can be used in the nano devices. 

The rest of this paper is organized as follows. In Sec.~\ref{Calculation_method}, the calculation methods are introduced. In Sec.~\ref{Results and Discussion}, the DFT results on the structural stability and the electronic structures of  Sc$_2$C(OH)$_2$ are discussed. To facilitate the understanding of DFT results, an image-potential well model is proposed and the corresponding Schr\"{o}dinger equation is solved. The effects of interlayer distances and external electric fields on the energetics of the NFE states and consequently the band gap tuning are demonstrated. Sec.~\ref{Conclusion} is the conclusions.

\section{Calculation methods}
\label{Calculation_method}

First-principles calculations were performed within the framework of DFT by the Vienna ab initio simulation package (VASP).~\cite{G. Kress, G. Kress2, G. Kress3, G. Kress4, G. Kress5} The exchange-correlation energy is approximated within the generalized gradient approximation proposed by Perdew, Burke, and Ernzerhof (GGA-PBE).~\cite{J. P. Perdew} A cutoff energy of 520~eV was used for the plane wave basis set.~\cite{M. Khazaei} All the atoms in the primitive cells (two Sc atoms, one C atom, two O atoms and two H atoms) were fully relaxed until the force acting on each atom was less than $10^{-4}$~eV/{\AA}. The convergence of the electronic properties was less than 10\textsuperscript{-5}~eV. A 21$\times$21$\times$1 Monkhorst-Pack (MP) \textit{k}-mesh was used to perform the geometrical relaxation calculations of the primitive cells and a 33$\times$33$\times$1 \textit{k}-mesh was employed for all energy band calculations.~\cite{H. J. Monkhorst, M. Methfessel} The  projector augmented wave (PAW) method was adopted.~\cite{G. Kress2} To investigate the responses of the NFE states to the external electric fields, uniform electric fields were applied along the $z$ axis perpendicular to the monolayer plane.~\cite{G. Makov, J. Neugebauer} To reduce the error in the energetics of states in the presence of the electric fields, the surface dipole were corrected because of the possible change in symmetry and the periodical boundary condition (PBC).~\cite{A. Natan, H. Komsa} Therefore, a planar dipole layer was introduced in the middle of the vacuum region to avoid interactions between the periodically repeated images.~\cite{A. Natan, J. Neugebauer} 

Though ScC$_x$(OH) and Sc$_2$CCl$_2$ exist experimentally,~\cite{J.Zhou2019, S.J.Hwu1986} as far as we know, the successful synthesis of Sc$_2$C(OH)$_2$ has not been achieved yet. To confirm its stability, the phonon spectra were calculated by Phonopy~\cite{A.Togo,K. Parlinski}. The force constants were extracted from the 4$\times$4$\times$1 supercells.~\cite{K. Parlinski}

The transport properties were calculated via the non-equilibrium Green's function (NEGF) scheme by Atomistix ToolKit (ATK).\cite{ATK,M. Brandbyge,J. M. Soler} The calculations were   performed using the GGA-PBE functional in conjugation with double-$\zeta$ polarization numerical basis sets. A 21$\times$21$\times$1 MP \textit{k}-mesh is also used to optimize the primitive cell with the force tolerance {0.01~eV/\AA} by ATK package. A cutoff energy of 500 Ry was used for density mesh. The self-consistent calculations were continued until the total energies were converged with $10^{-5}$ Ry tolerance. To calculate the transport properties, the \textit{k}-point mesh was 1 $\times$10$\times$60 in the first Brillouin zone. The current through the device was obtained at finite bias voltages using the Landauer formula\cite{S. Datta}. The gate fields were applied along the $z$ direction. Due to the extensive distribution of the wave functions of NFE states outside the surface, the tail of NFE state wavefucntions are not produced perfectly using the conventional localized basis sets. Hence, three layers of ghost atoms were added on each side of the layer to reproduce the energy bands in the presence of external electric field correctly (shown in Fig. S9 of the Supporting Information).~\cite{NFE} In ATK, a ghost atom represents a point in the space with a localized basis set but without any pseudopotential core or real atomic nucleus.~\cite{NFE} The separation between the neighboring ghost atom layers is 1.0~\AA. The introduction of the ghost atom layers significantly improve the agreements between the computed band structures using ATK with localized basis sets and those obtained from VASP with plane wave basis sets  as shown in Fig. S9.

\section{Results and Discussion}
\label{Results and Discussion}
\subsection{Structural Information}
\label{Geometries}

When a hydroxyl group is absorbed on the surface of Sc$_2$C monolayer, three different sites are available. With the definitions in Ref.~[67], these sites are named as A, B, and T,  and shown in Fig.~\ref{structure_OH}.~\cite{H. Weng} Site T is on the top of the Sc atom; site A is the hollow site above the C atom; and site B is the hollow site above the Sc atom of the opposite surface. The definitions of the hollow positions have the difference from the definitions in Ref.~[13]. Regarding the combinations of these three absorption sites on two surfaces, six distinct configurations can be constructed.~\cite{M. Naguib, M. Khazaei, H. Weng} 
 
 \begin{table}[tb]
\caption{
The total energies (in $eV$ per unit cell) of six possible configurations of Sc$_2$C(OH)$_2$. The name of models indicate the sites (A, B, T shown in Fig.~\ref{structure_OH}) where the OH groups were adsorbed. The energy of the most favorable structure (BB) is set to zero and the energies of other structures are relative to it. 
}\label{stablestructure}
\begin{tabular*}{0.5\textwidth}{@{\extracolsep{\fill}}c|cccccc}
 	\hline
 	\hline
sites of OH & BB & BA & AA & TB & TA & TT\\
         \hline
Sc$_2$C(OH)$_2$ & 0.000 & 0.371 & 0.770 & 1.384 & 1.794 & 2.669 \\
 	\hline
 	\hline
\end{tabular*}
\end{table}

To find the most energy favorable stable structure, all atoms and their lattice constants in the primitive cells of these six possible structures are fully relaxed. Their symmetries are taken into consideration before the structures are relaxed. For example, the space group of BB-type structure is $P\overline{3}m1$. Their energies after relaxations are summarized in Table~\ref{stablestructure}. Here, BB-type structure is the reference structure and the energies of other structures are the relative values to it. The calculations reveal that BB model is the most energy favorable structure. Since the successful synthesis of the functionalized Sc$_2$C(OH)$_2$ monolayer has not been achieved yet, the stabilities of all model structures are investigated using the phonon spectra calculations. It is found that only BB, AB and AA structures are dynamically stable because their vibrational frequencies are all positive. It is also observed that when one or both OH groups are adsorbed on T site(s), the models are energetically unstable. In other words, the OH groups are favorably adsorbed on the hollow sites. This result is consistent with the previous results as reported by M. Khazaei \textit{et al.} and H. Weng \textit{et al.}~\cite{M. Khazaei,H. Weng} 

The structures of these stable structures are shown in Fig.~\ref{structure_OH} and their phonon spectra can be found in Fig. \ref{phonon}. For the sake of visibility of the acoustic bands, two highest phonon bands, resulting from the O-H stretching modes are not shown. The climbing image nudged elastic band (cNEB) method is also applied to find the reaction path connecting these structures.~\cite{G. Henkelman,G. Henkelman2} The result in Fig.~\ref{phonon}(a) indicates that these structures are local minima in energy, and separated by energy barriers. The cNEB calculation provides another evidence to confirm the stabilities of these structures.

BB-type structure is the most stable structure, shown in Fig~\ref{structure_OH}(a). It has a hexagonal lattice with lattice parameters of \textit{a} = \textit{b} = {2.29~\AA}. In the calculations, the vacuum size is {30~\AA}. However, if a larger vacuum size ({60~\AA}) is adopted, the relaxed structure has only slight change. The C atom is the origin of the primitive cell. Its Wyckoff position is 1$a$, whose site symmetry is $D_{3d}$. Due to the crystal fields, the $p$ orbitals of C atom split into two groups.  The $p_x$ and $p_y$ orbitals form two-dimensional $E_u$ representation, whereas the $p_z$ orbitals form $A_{2u}$ one. The site symmetries of Sc atoms are $C_{3v}$. Five-fold degenerate $d$ orbitals are reduced into three sets: ($d_{x^2-y^2}$, $d_{xy}$) and ($d_{xz}$, $d_{yz}$) form two-fold degenerate $E$ representations and $d_{z^2}$ forms the one-dimensional $A_1$ one. The representation of $p$ orbitals split into two irreducible representations. $p_x$ and $p_y$ orbitals form $E$ representation and $p_z$ orbital forms $A_1$ one.  Because of the inversion symmetry operation, the symmetrical or anti-symmetrical combinations of Sc atomic orbitals form at $\Gamma$ point. 

\subsection{Monolayer Properties without electric fields}
\label{Properties without electric Fields}

To study the electronic structures of monolayers, a very large vacuum ({120~\AA}) is used to minimize the interactions between the neighboring layers in our DFT calculations. The fat bands of BB, AB, and AA Sc$_2$C(OH)$_2$ structures are plotted in Fig.~\ref{fat_bands}. The projected weights are achieved by projecting the wave functions onto the atomic orbitals (including $s$, $p$, and $d$ orbitals) of all atoms (Sc, C, O and H atoms) in the primitive cell, and the sizes of the dots represent the total-summation of these projected weights. It is found that BB, AB, and AA Sc$_2$C(OH)$_2$ are semiconductors with band gaps of about 0.45, 0.74, and 0.94~eV, respectively. AA (AB and BB) is (are) indirect (direct) band gap semiconductor(s).

Since the BB-type structure is the energetically favorable structure, its electronic properties are investigated in details. Compared the present energy bands with other reported results, for example, the energy bands reported by M. Khazaei \textit{et al.}, it can be found that their valence bands are very similar.~\cite{M. Khazaei} However, the unoccupied bands at high energies look different. This is because in the present calculations, a larger vacuum was used. Hence, more unoccupied NFE states with parabolic shape are emmerged. 

The valence bands are mainly composed of the atomic orbitals. To investigate them, the density of states (DOS) is decomposed and projected onto different atomic orbitals. The fat bands of the valence bands around $\Gamma$-point are also plotted in Fig.~\ref{pdos}. It is observed that at $\Gamma$ point, the valence band maxima (VBM) are two-fold degenerate. From Fig.~\ref{pdos}(d), it is seen that the Sc and C atoms have crucial contributions to the valence bands at $\Gamma$ point. The VBM wave functions are decomposed to the irreducible representation and projected onto different atoms. We found that they come mainly from the $d_{x^2-y^2}$, $d_{xy}$ orbitals of Sc atoms (shown in Fig.~\ref{pdos}(a)) and $p_x$, $p_y$ orbitals of C atom (shown in Fig.~\ref{pdos}(c)). These orbitals form the $E_u$ representation at $\Gamma$ point.

The conduction bands possess parabolic energy dispersions, which are analogous to the free electron states. These bands do not come from the atomic orbitals because their projection weights are extremely low, as shown in Fig.~\ref{fat_bands}. These bands are NFE bands, which result from the IP states.~\cite{NFE,M. Feng, M. Feng2} To investigate the spatial distribution of the NFE states of BB-Sc$_2$C(OH)$_2$, the decomposed charge densities of the first six NFE bands at $\Gamma$ point are plotted in Fig.~\ref{density}. Here, the decomposed charge densities have been averaged along the $z$ axis. The corresponding side views of the decomposed charge densities are below them. Since BB-Sc$_2$C(OH)$_2$ possesses inversion symmetry, "$+$"/"$-$" is used to indicate the even/odd parity of the wave functions. In Fig.~\ref{density}(a), the $1^+$ and $2^-$ NFE bands are shown. As seen, their averaged charge densities peak several angstroms away from the outmost H atoms and extend in the vacuum region. The outer layer distribution of the NFE states provide remarkable transport channels for the electrons to pass through without being scattered by thermal fluctuation of the atoms on the surfaces.~\cite{J. Zhao charging} As a consequence, if the NFE states can be modulated, the Sc$_2$C(OH)$_2$ monolayer can be a promising candidate for the electronic device applications. 

\subsection{Image-potential well model}
\label{image-Potential}

In our previous DFT calculations, we revealed that the OH-terminated MXene monolayers obtain ultra low work functions attributed to the polarity of the hydroxyl groups.~\cite{OH} Moreover, we anticipated that the NFE states are energetically found close to the Fermi energy because of the shallow potential tails above the surfaces of the OH-terminated MXenes.~\cite{NFE} In DFT calculations, the potentials in the bulk and in the short range of the surface are fairly accurate. However, DFT fails to reproduce the Coulomb-like tails of the surface potentials in the long range.~\cite{V. M. Silkin} Here, to shed light on the formation of the NFE states above the Sc$_2$C(OH)$_2$ monolayer, we have adopted a model to simplify the one-dimensional image-potential well. We address the important role of the Coulomb-like potential tail played in the formation of the NFE states. 

The potential well is defined as: 
\begin{equation}
V(z)=\begin{cases}
-V_0 & |z| \leq a\\
-\frac{e^2}{4(|z|-a+\Delta)}  &|z| > a
\end{cases}
\end{equation}
where $\Delta=\frac{e^2}{4V_0}$. This potential well can be divided into two parts: 1) within the monolayer, the potential well has a flat bottom whose depth is $V_0$ because the electrostatic potential is averaged. The width of the monolayer is $2a$.~\cite{NFE}  2) In the vacuum region, a Coulomb-like image-potential tail is adopted.~\cite{V. M. Silkin}
Compared with the potential calculated from DFT, the dipole moment on the surface and the corrugation of the surface are not taken into consideration.~\cite{NFE,M. Weinert} Therefore, this image-potential is a simplified model and highlights only the Coulomb-like potential tail. The width of the Sc$_2$C(OH)$_2$ monolayer is the distance between the outmost H atoms in the primitive cell ($a \simeq 3.5 \AA$). By fitting the model potential to the DFT potential results, the depth of the well is estimated to be $V_0\simeq$ 9.5 eV. By solving the Schr\"{o}dinger equation for the model potential, the bound solutions can be found in Supporting Information.

In principle, the number of solved solutions is infinite. Dependent on the behavior of the wave functions in the vacuum region, the solutions can be categorized into two groups: 1) As shown in Fig.~\ref{WaveNFE}(a), the lowest four eigenstates decay to zero exponentially when leaving away from the surfaces into the vacuum. These surface states are ascribed to Shockley states because of their rapid change in the electron potential associated solely with the surface termination.~\cite{W. Shockley} To address this issue, their counterparts of a square well with the same width and depth are given in Fig.~\ref{WaveNFE}(c). 2) Other solutions, as plotted in Fig.~\ref{WaveNFE}(b), differ from the Shockley states significantly because their charge density maxima are in the vacuum region. The corresponding wave functions are much more extensive along the $z$ axis. They are the NFE states and result from the Coulomb potential tail. For the monolayer, these eigenvalues with larger principal quantum number $n$ ($n>4$) form the Rydberg series, as shown in Fig.~\ref{Rydberg}. We found that the eigenvalues $E_n$ obey the following formula:
\begin{equation}\label{energy}
E_n=-\frac{3.4217 eV}{(n-3.3703)^2}\simeq-\frac{E_0}{4(n-3.3703)^2}
\end{equation}
where $E_0 = 13.6~eV$. If the principal quantum number $n$ is very large, the eigenvalues reads $E_n=-0.85/n^2$, which is the same as the Rydberg series of a conducting metal surface.

By comparing the DFT results with the analytical solutions, some differences can be identified. For example, the principal quantum number $n$ in Eq.~\ref{energy} can be any integer. However, DFT calculations show that the number of the bound NFE states is small, and depends on the size of the vacuum region as plotted in Fig.~\ref{bands}. The valence bands in Fig.~\ref{bands} are independent from the size of the vacuum region. The energetics of the unoccupied bands change significantly with respect to the vacuum size. For example, as the size of the vacuum gets larger, more parabolic bands appear at $\Gamma$ point, but it has no effect on the energy gap. These bands represent the NFE states. When the vacuum size is small, as shown in Fig.~\ref{bands}(a), it affects the energy gap. In other words, the band gap is sensitive to the vacuum size. As the vacuum size gets decreasing, the band gap becomes smaller and smaller. The band gap width evolution with respect to the interlayer distances is found in Table~S2 of the supporting Information illustrating that when the interlayer distance is less than 20{~\AA}, the band gap width becomes sensitive to it. Another difference comes from the asymptotic behavior of the wave functions of the NFE states. The general analytical solutions of the image-potential model in the Supporting Information indicate that the wave functions of NFE states are extensive along the $z$ axis but approach to zero gradually when they are far away from the surfaces, as shown in Fig.~\ref{WaveNFE}(b). The charge densities of two lowest NFE states from DFT calculations in Fig.~\ref{density}(a) imply that they follow such behavior. Although a very large vacuum size ({120~\AA}) is adopted in the calculations, it is failed to find that the wave functions of NFE states with higher energies such as $3^+$ in Fig.~\ref{density}(b) and $5^+$ in Fig.~\ref{density}(c), approach to zero when they are away from the surfaces.

In the previous DFT calculations, GGA-PBE functional was used. As it is known to all, GGA-PBE functional is valid only for the short range and the predicted band gap width is typically underestimated.~\cite{J. Perdew} To improve the accuracy and examine the simulation results, the long-range correction procedures, such as Heyd-Scuseria-Ernzerhof (HSE06) screened hybrid and its parent PBEh (also called PEB1PBE or PBE0) global hybrid, are performed.~\cite{T. M. Henderson, M. Ernzerhof, C. Adamo} A less dense 13$\times$13$\times$1 MP \textit{k}-mesh is adopted in these time-consuming calculations. As shown in Table~S1 of the Supporting Information, after relaxations, BB-structure is the most energy favorable structure as well. The corresponding energy bands with respect to different vacuum sizes can be found in the Supporting Information (Fig.~S1 and Fig.~S2). Compared with the GGA-PBE results, it can be noticed that the energy band widths are much larger. For example, in the vacuum size of {20~\AA}, the band gap width obtained from GGA-PBE scheme is about 0.45~eV while HSE06 gives 0.80~eV and it becomes 1.39~eV in PBEh calculation. HSE06 calculations include the screening effect over a long range (roughly in the order of two or three chemical bonds). Typically, the results obtained from HSE06 calculations are more accurate in the extended systems than those from PBEh. This is because of the extensive distribution of the charge density in solids inducing long-range exchange and correlation interactions.~\cite{T. M. Henderson, A. F. Izmaylov} These two interactions have the opposite signs and their cancellations lead to the screening effect.~\cite{T. M. Henderson} In the HSE06 functional, this screening effect is included while in the PBEh functional, the correction of the screening effect is absent.~\cite{C. Adamo}

\subsection{Vacuum size dependent properties}
\label{Vacuum size dependent properties}

The above differences can be attributed to the artificial supercells used in the first-principles calculations when the PBC is applied. Generally, in order to simulate various properties of a two-dimensional material accurately, it is enough to adopt a large vacuum region to reduce the interlayer overlap of wave functions and the interaction. However, when we deal with NFE states, the properties related to the NFE states might be affected even by employing a large vacuum region. As plotted in Fig.~\ref{WaveNFE}(b), when the principal quantum number $n$ increases, its corresponding NFE wave functions become very extensive spatially. Sometimes, the overlap of these extensive wave functions from the neighboring layers is significant, and their hybridization becomes important. The asymptotic behaviors of the wave functions with large principal quantum number $n$ also change due to the superposition of the Coulomb-like potential tail from the neighboring monolayer in the vacuum region. Therefore, the artificial images imposed by the PBCs results in the overlap of the extensive wave functions of NFE states. DFT fails to reproduce the conduction bands representing NFE states of monolayer accurately because the influence from the neighboring layers cannot be neglected. Basically, the DFT results always represent the electronic structures of the NFE states in multiple layers not for the monolayers.

To analytically understand the above results, we study the NFE states by solving the one-dimensional Schr\"{o}dinger equation in the presence of a neighboring monolayer. With the superposition of the electrostatic interaction from the nearest neighboring monolayer,  the potential along the $z$ axis becomes,
\begin{equation}\label{Potential well model} 
V(z)=\begin{cases}
-V_0 & z \geq L-a\\
-\frac{1}{4(z+L-a+\Delta)}-\frac{1}{4(-z+L-a+\Delta)}+\frac{1}{4(-2a+2L+\Delta)} &|z|<L-a\\
-V_0 & z \geq-L+a
\end{cases}
\end{equation}
where, $2L$ represents the distance between the centers of the neighboring layers ($-L\leq z \leq L$). To compare with the previous DFT calculations, $L$ is {60~\AA}. Since we are interested in the NFE states at $\Gamma$ point, the Bloch wave function is simplified to $\phi(z)=e^{ik_{\Gamma}z} u(z)=u(z)$. The eigenvalues and wave functions of bound states are solved numerically by the shooting method.~\cite{S. E. Koonin} The charge densities of all NFE bound states are shown in Fig.~\ref{densNFE}.

In Fig.~\ref{densNFE},  the NFE states are categorized into two groups: 1) As shown in Fig.~\ref{densNFE}(a), some of the NFE states behave similar to those obtained for an individual monolayer. For example, the density of the $1^+$ state approaches to zero when it is about {15~\AA} away from the center of the monolayer. 2) However, as shown in Fig.~\ref{densNFE}(b),  there exist some NFE states which are much more extensive and are totally distinct from the solutions in monolayer case. These solutions facilitate us to understand the results of DFT calculations. Compared with the results shown in Fig.~\ref{density}, it can be easily found that they are very similar. For example, their charge density maxima are in the vacuum region. The number of the  NFE bound states is finite. Both DFT results and the numerical solutions have parities. For some states with even parities, for examples $5^+$ and $7^+$ in Fig.~\ref{densNFE}(b), the density maxima appear in the middle of the vacuum region between the layers. 

The hybridization can account for these distinctions. For the $1^+$ state, the interlayer overlap is extremely small, therefore, its wave function and density distribution have less change. However, in Fig.~\ref{densNFE}(b), these states result from the higher principal number $n$. They are very extensive and can be hybridized significantly. The $5^+$ and $7^+$ states are analogous to the bonding state of a H$_2$ molecule and the $6^-$ state is similar to the anti-bonding state. The decreasing of vacuum size facilitates the overlap, and if the vacuum size is small enough, the overlap between the lowest energy NFE states is possible. As indicated in Fig.~\ref{bands}, if the vacuum becomes less than {20~\AA}, the band gap width gets smaller due to the overlap and hybridization.

Experimentally, it has been reported that the interlayer distances between monolayers could be varied in the range of 10 to {120~\AA}.~\cite{Y. Ma} The band gap width dependence on the vacuum size suggests that it is possible to engineer the band gap width by controlling the interlayer distances of functionalized MXene materials and fabricate the devices with different band gap widths. 

\subsection{Effect of electric field on the NFE states}
\label{Properties with electric Fields}

As shown in Fig.~\ref{gap_mod}, the NFE states respond to the external electric fields resulting in the modulation of energy gap widths.~\cite{M. Ishigami,K. H. Khoo} Generally, it is found that by increasing the magnitude of the electric field, the energy gap width becomes smaller and smaller, and a semiconductor can be converted to a metallic system. However, the response of the band gap width with respect to the applied electric fields is also vacuum size dependent. If the vacuum size is small, for example {15~\AA}, a much larger electric field is required to convert Sc$_2$C(OH)$_2$ from semiconducting to metallic. However, if the vacuum size is large, the response changes abruptly. For example, with the vacuum size of {50~\AA} and the electric field of about 0.08~eV/\AA, the band gap width reduces suddenly.

The energy band responses of Sc$_2$C(OH)$_2$ with different interlayer distances to the external electric fields can be found in the Supporting Information. It is found that the energy bands representing NFE states are sensitive to the fields. The band gap widths can be reduced with the increase of fields. Therefore, upon increasing the field, the Sc$_2$C(OH)$_2$ turns into conductor and the band gap is closed. Using HSE06 and PBEh, the responses of the energy bands with respect to the external electric fields are investigated. The interlayer distance is {20~\AA} and the results are shown in Fig.~S8. Compared with previous results from GGA-PBE as shown in Fig.~\ref{gap_mod}, HSE06 calculations show that {0.32~V/\AA} is required to convert the semiconducting Sc$_2$C(OH)$_2$ into a conductor. This value is close to the result from GGA-PBE calculation. However, a much higher field ({0.45 V/\AA}) is required to close the gap of the band structures obtained from the PBEh scheme due to the existing larger band gap width (1.39 eV).

Perturbation theory can be used to understand the above observations. The perturbation term is
\begin{equation}
H'(z)=-eEz
\end{equation}
where $E$ is the magnitude of the applied electric field along the $z$ axis and $-e$ is the charge of electron. The energy shift is
\begin{equation}\label{perturbation}
\Delta_n=E_n-E_n^0=H'_{nn}+\sum_{k\neq n}\frac{H'^2_{nk}}{E_n^0-E_k^0}
\end{equation}
where $H'_{nk}=-eE\langle n|z|k\rangle$ and $|k\rangle$ is the unperturbed eigenstate.
At $\Gamma$ point, the representation of the perturbation term is $A_{2u}$. Given the fact that the eigenstates have even or odd parities with respect to the $z$ axis, while the perturbation term $H'$ is odd, the first order energy shift term $H'_{nn}$ is zero. Therefore, the second order term is significant. The VBM is two-fold degenerate ($E_u$) at $\Gamma$ point and its degeneracy can not be lifted in the presence of electric fields ($A_{2u}$). The band gap width is mainly determined by the NFE states because CBM originates from them. 

In Eq.~\ref{perturbation}, the energy shift is proportional to $E^2$. It fits well when the field magnitude is small. The energy shift behaves linearly with respect to $\langle n|z|k\rangle^2$, which depends on the extension of the wave functions in the $z$ direction. From Fig.~\ref{WaveNFE}(b) and Fig.~\ref{densNFE}, it is observed that the NFE states with higher energies have more extensive wave functions. This explains the sudden drop of gap widths when the vacuum size is large. As the vacuum size increases,  more bound NFE states are found in Fig.~\ref{bands}, and they are much more extensive than the lowest two NFE states. This makes more contribution to the energy shift term. The figures in the Supporting Information indicate that in the presence of large external fields, the NFE states with higher energies are much more sensitive to the field than those with lower energies. They exchange their positions resulting in the sudden reduction of band gap widths.  

\subsection{Transport properties}
\label{Transport properties}
The NFE states can be identified from the other bands, which results from the atomic orbitals, because of their unique spatial distributions outside of the sursafce. Moreover, in the presence of external fields, the semiconducting OH-terminated MXenes turn into conducting ones. To investigate whether the NFE states can contribute to the electron transport, NEGF is used to calculate the \textit{I-V} curves as a function of gate field.

The relaxed primitive cell of BB-type structure from VASP calculations is used as the initial structure, and it is fully optimized by ATK before performing the electron transport calculations. As expected, the optimized structure is very similar to the initial one. For example, the change in the lattice constants is less than {0.03~\AA}. There is no change in the space group. Since long-range Coulomb-tail is involved and the wave functions of NFE states reach their density maxima in the vacuum region, the numerical atomic orbitals basis sets centering the positions of atoms fail to achieve the NFE bands accurately. 

To study the manipulation of the transport properties, it is required to reproduce the gap widths response by ATK. Since the wave functions of the NFE states do not come from the atoms in the primitive cell, three layers of ghost atoms are added on each surface of the configuration to better representation of the NFE states outside the surface.~\cite{NFE} Double-$\zeta$ polarization numerical basis sets are used for these ghost atoms. The configuration and the band gap width response to the external electric fields are shown in Fig.~S8(b) of the Supporting Information. After the introduction of the ghost atoms, it is found that the response of the band gap width from ATK is very similar to that from VASP calculations. For example, in the absence of the field, the band gap width is about 0.48~eV, which is very close to the result (0.45~eV) from VASP. The band gap widths are reduced upon increasing the magnitudes of the electric fields. At an external electric field of about 0.29~V/{\AA}, Sc$_2$C(OH)$_2$ turns metallic. These facts provide the evidence for the necessity of adopting the ghost atom scheme in the electron transport simulations of nanodevices with NEF states.

In the calculations, the distance with the neighboring monolayers is {20~\AA}.  As shown in the inset of Fig.~\ref{IV}, the gate field is applied in the $z$ direction and Sc$_2$C(OH)$_2$ is connected by two electrodes. In the absence of the gate field, the current is extremely small when the source-drain voltage is applied. This is reasonable because the energy gap is about 0.4 eV. If the voltage between two electrodes is too small, the electron fails to get enough energy to overcome the energy gap of Sc$_2$C(OH)$_2$. Therefore, only few electrons transport from one electrode to the other and the current is small. By increasing the gate field, the band gap is reduced. For example, if the gate field is about {0.2~V/\AA}, the current increases when the source-drain voltage is about 0.2~V. By further increasing the gate field, the energy gap is closed and the semiconducting Sc$_2$C(OH)$_2$ converts into a conductor, whose current increases significantly. The \textit{I-V}  curves in Fig.~\ref{IV} indicate that the NFE states can affect the transport current. Hence, Sc$_2$C(OH)$_2$ can be one of the suitable candidates for nano switching applications because of its current modulation by the gate field.

\section{Conclusions}
\label{Conclusion}
The interesting electronic structures of hydroxyl-functionalized MXenes has been the subject of many theoretical and experimental studies.\cite{NFE,OH, Q.Peng2014, S. Kumar2016, E.Balci2018, J.Yang2016} Here, we highlight the influence of the NFE states on the electronic structure of the Sc$_2$C(OH)$_2$. Based on the first-principles calculations and the image-potential well model, it is found that NFE sates have very extensive wave functions in the vacuum regions with the neighboring layers. The NFE states of two neighboring monolayers can overlap and be hybridized. We propose that the energy gap width of Sc$_2$C(OH)$_2$ can be engineered by the interlayer distance. For example, if the interlayer distance is less than {20~\AA}, the band gap width is reduced with the decrease of the interlayer distance and a higher gate field is required to convert it into a conductor. If the distance is about {30~\AA}, the band gap width becomes insensitive to the change of the interlayer distance and most fundamental properties of the monolayer can be predicted with high accuracy. At Larger distance ($>$ 60~\AA), more NFE states with higher energies can be observed. The general trends and physics of the NFE states are independent from the type of the employed functionals (here, PBE, PBEPBEh and HSE06) in the DFT calculations. In the presence of a finite gate field with the magnitude of {0.3~V/\AA}, Sc$_2$C(OH)$_2$ can be converted into a conductor and the NFE states can be used as the channels for the electrons to transport.  These properties indicate that hydroxyl-functionalized MXenes, such as Sc$_2$C(OH)$_2$, are the prospective candidates for the nano switch devices or sensors.

\bigskip

\begin{acknowledgements}
We would like to express sincere thanks to the crews in the Center for Computational Materials Science in the Institute for Materials Research (Tohoku University) for the support.

\end{acknowledgements}

\newpage

\begin{flushleft}
{\large \textbf{Captions} }
\end{flushleft}

\begin{figure}[tbp]
\includegraphics[width=1\textwidth]{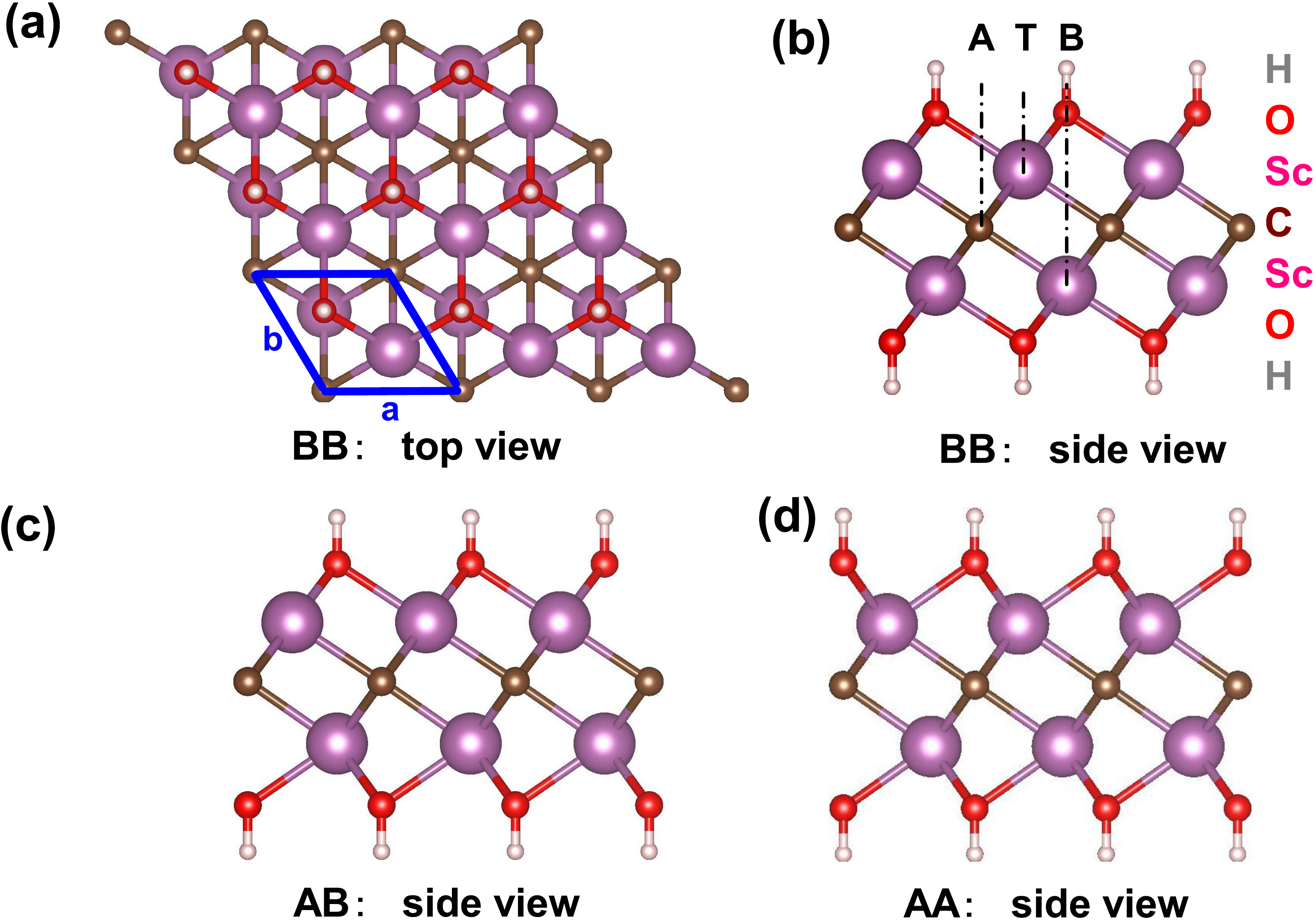}
\caption{J. Zhou \emph{et al.}} \label{structure_OH}
\end{figure}

\begin{figure}[tbp]
\includegraphics[width=1\textwidth]{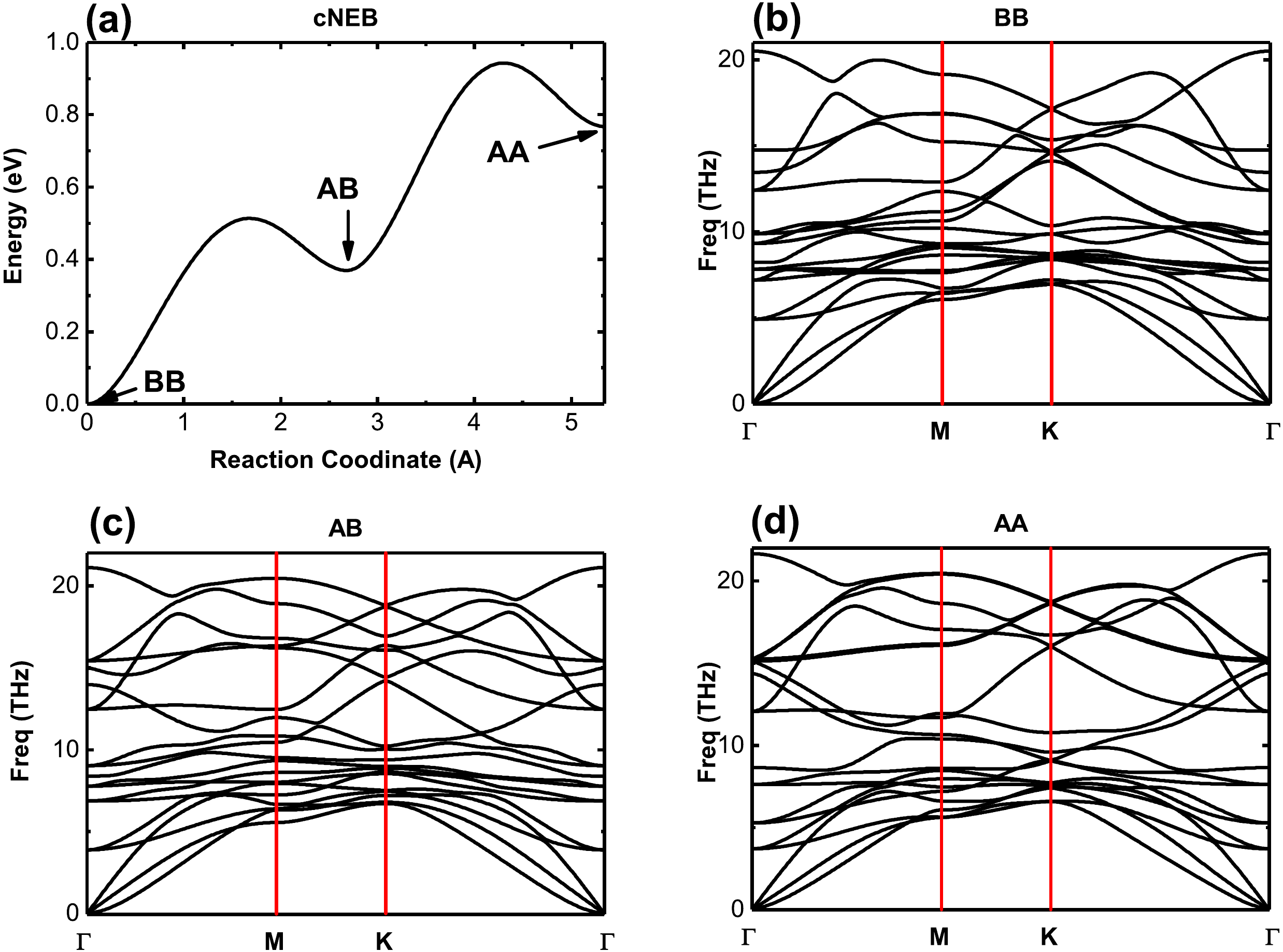}
\caption{J. Zhou \emph{et al.}} \label{phonon}
\end{figure}

\begin{figure}[tbp]
\includegraphics[width=1\textwidth]{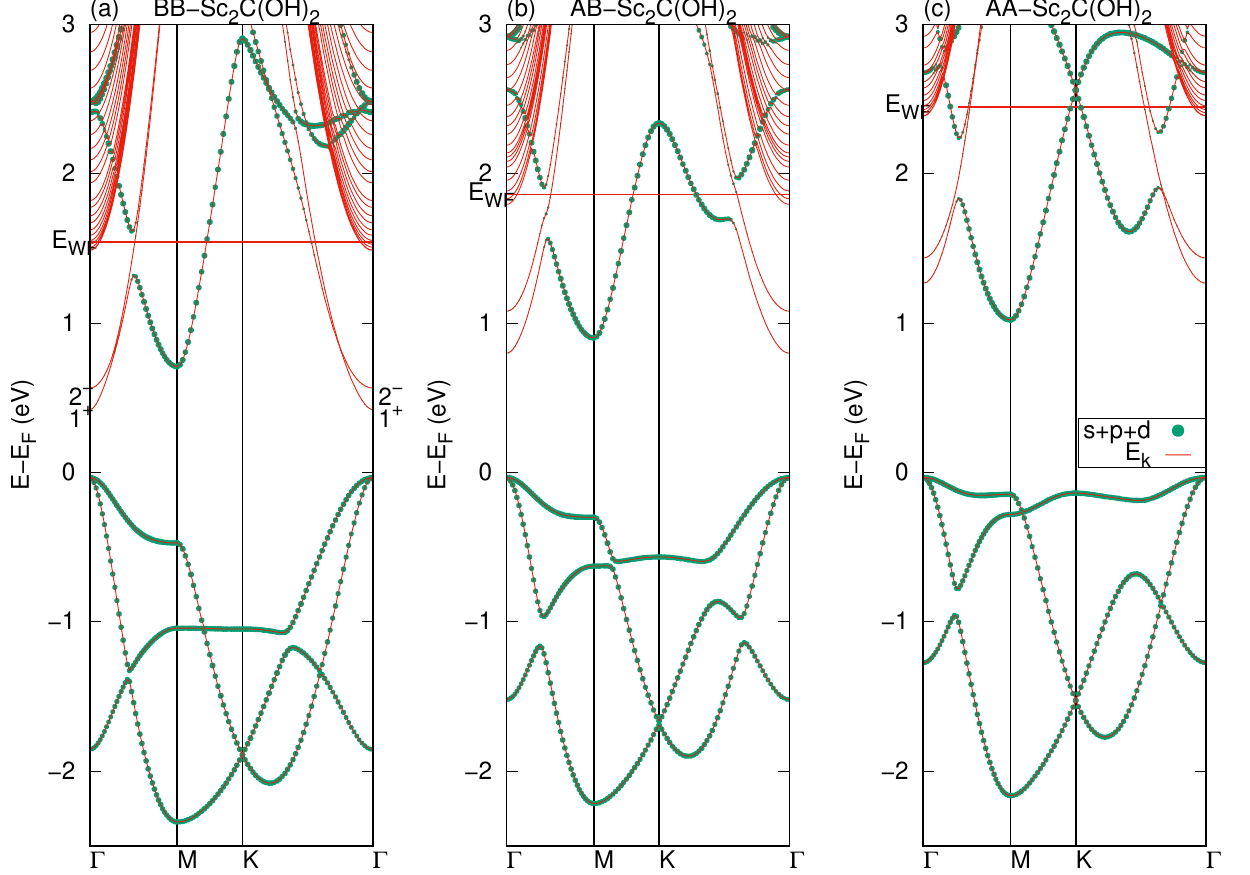}
\caption{J. Zhou \emph{et al.}} \label{fat_bands}
\end{figure}

\begin{figure}[tbp]
\includegraphics[width=1\textwidth]{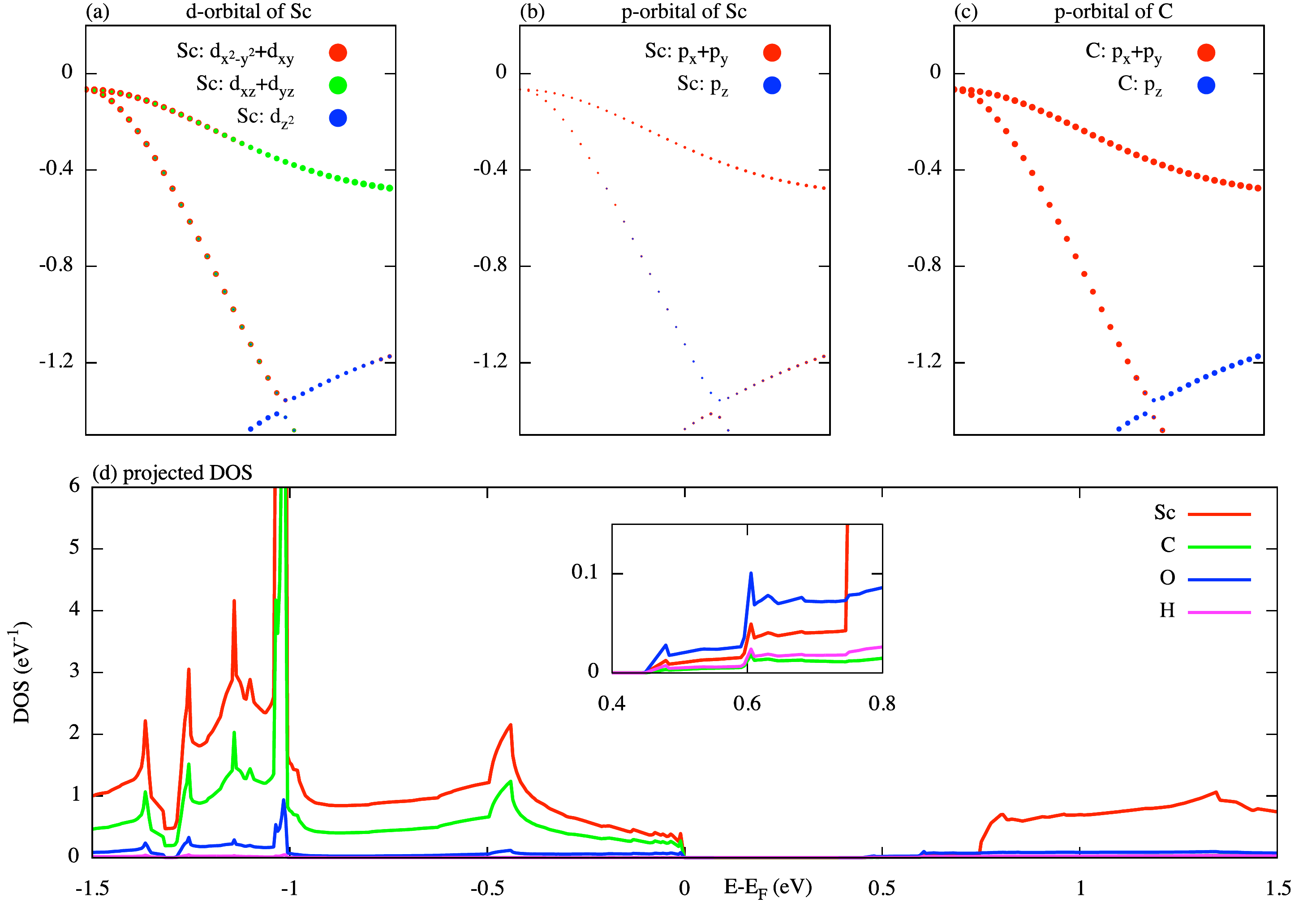}
\caption{J. Zhou \emph{et al.}} \label{pdos}
\end{figure}

\begin{figure}[tbp]
\includegraphics[width=1\textwidth]{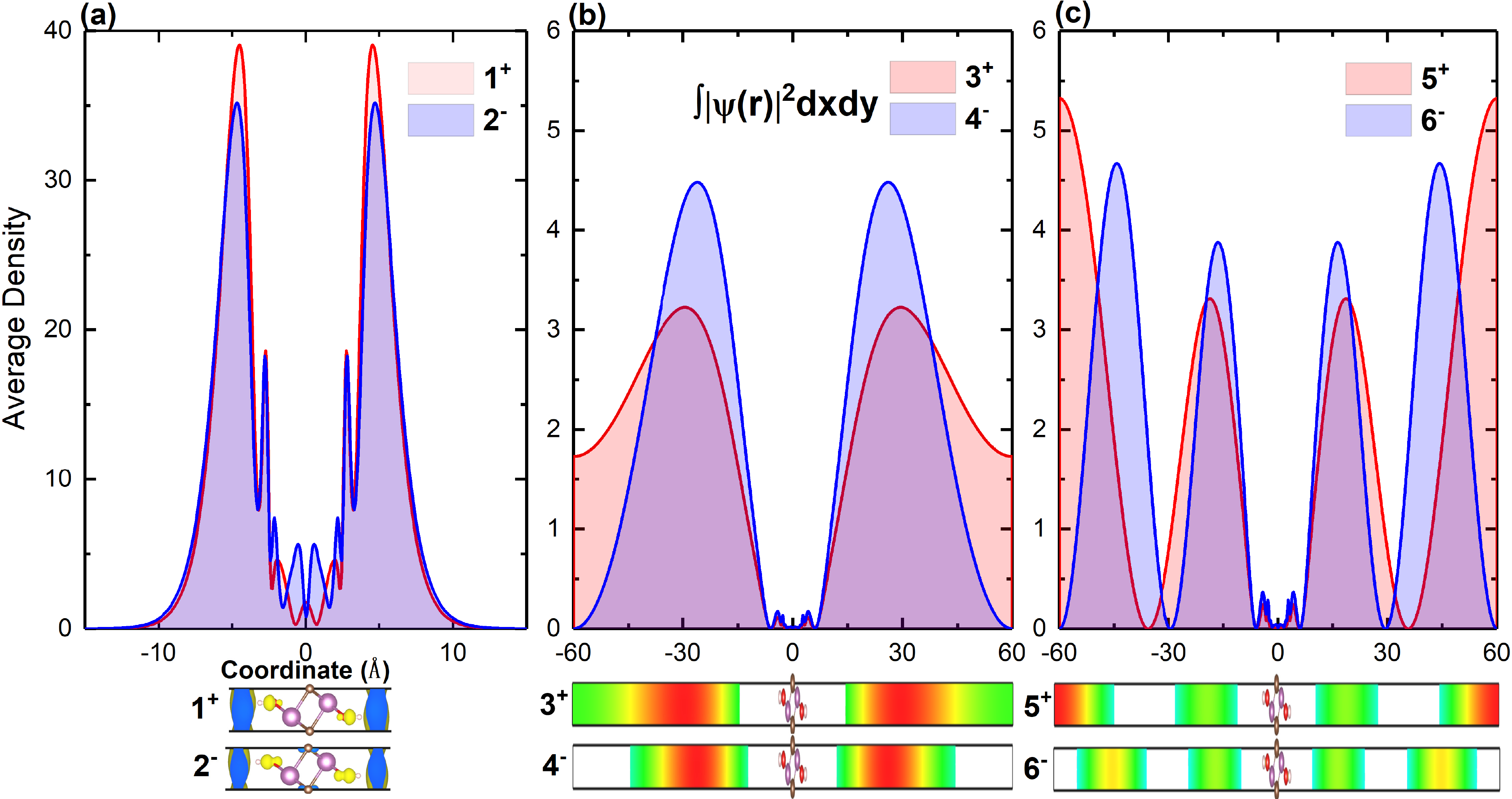}
\caption{J. Zhou \emph{et al.}} \label{density}
\end{figure}

\begin{figure}[tbp]
\includegraphics[width=1\textwidth]{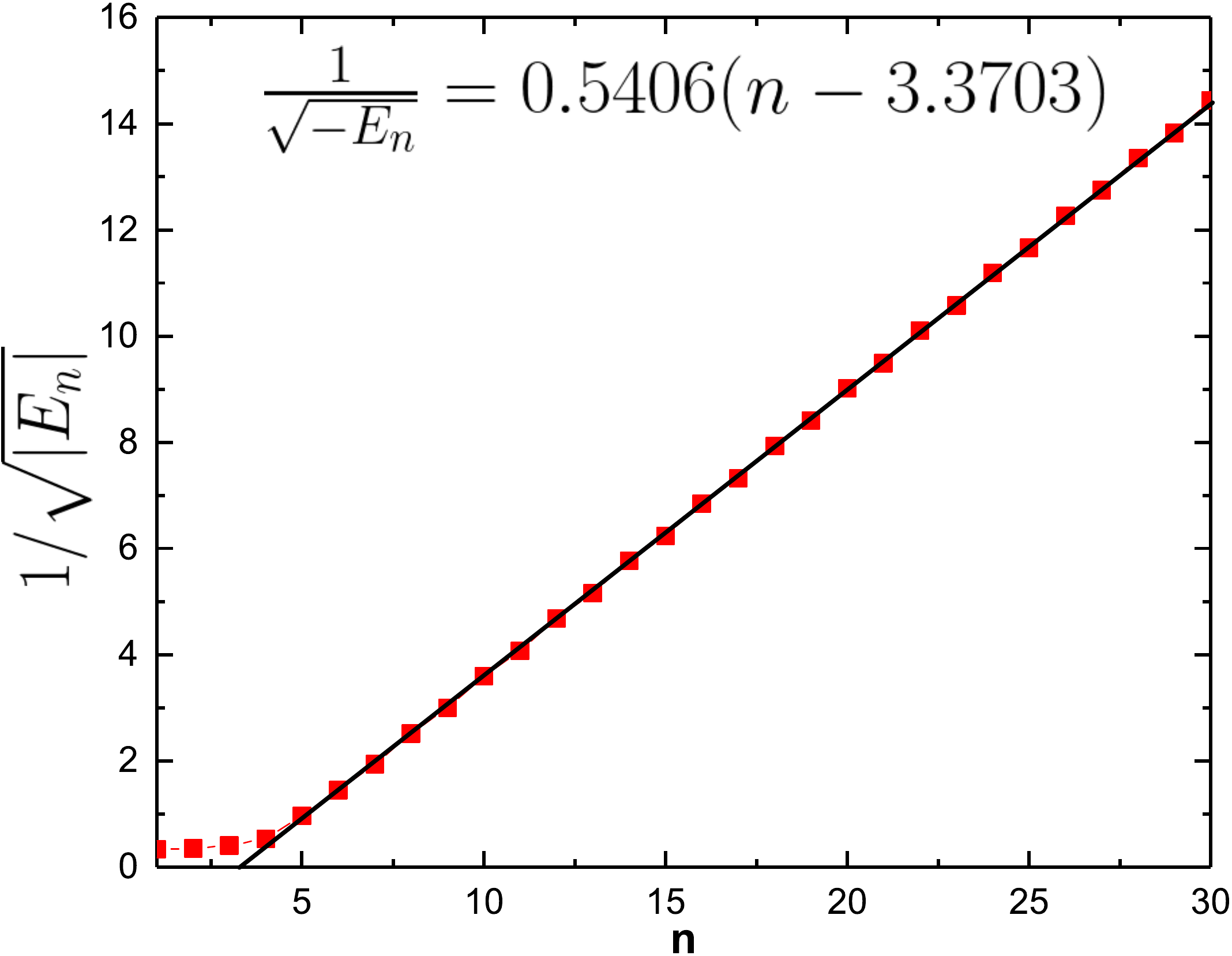}
\caption{J. Zhou \emph{et al.}} \label{Rydberg}
\end{figure}

\begin{figure}[tbp]
\includegraphics[width=1\textwidth]{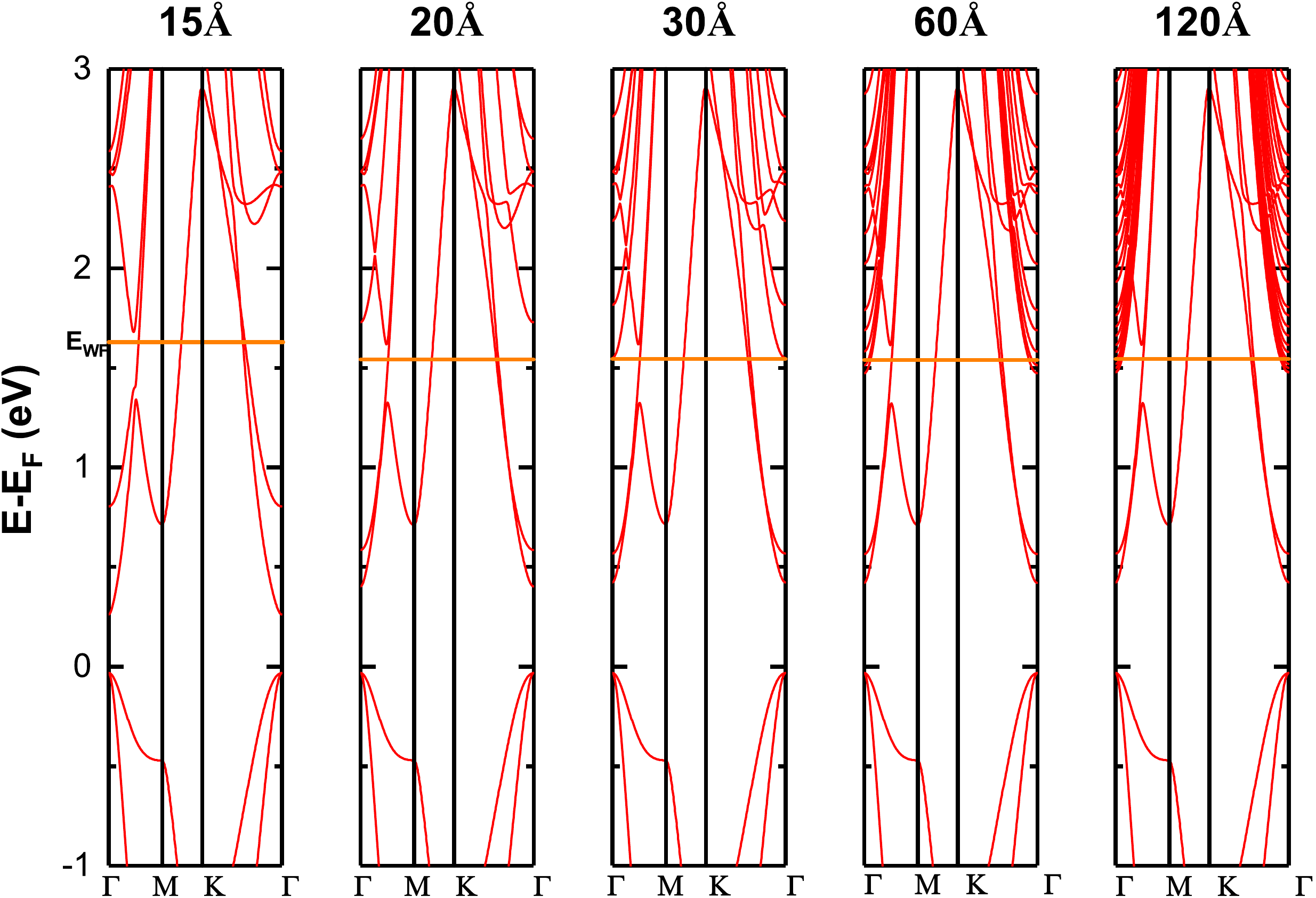}
\caption{J. Zhou \emph{et al.}} \label{bands}
\end{figure}

\begin{figure}[tbp]
\includegraphics[width=1\textwidth]{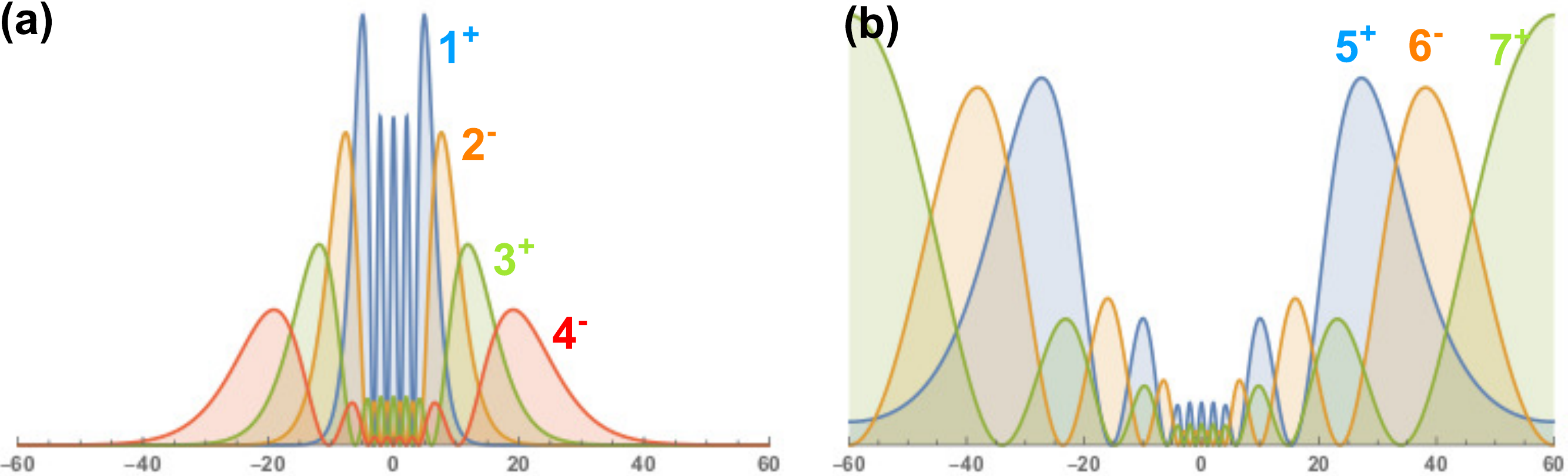}
\caption{J. Zhou \emph{et al.}} \label{densNFE}
\end{figure}

\begin{figure}[tbp]
\includegraphics[width=1\textwidth]{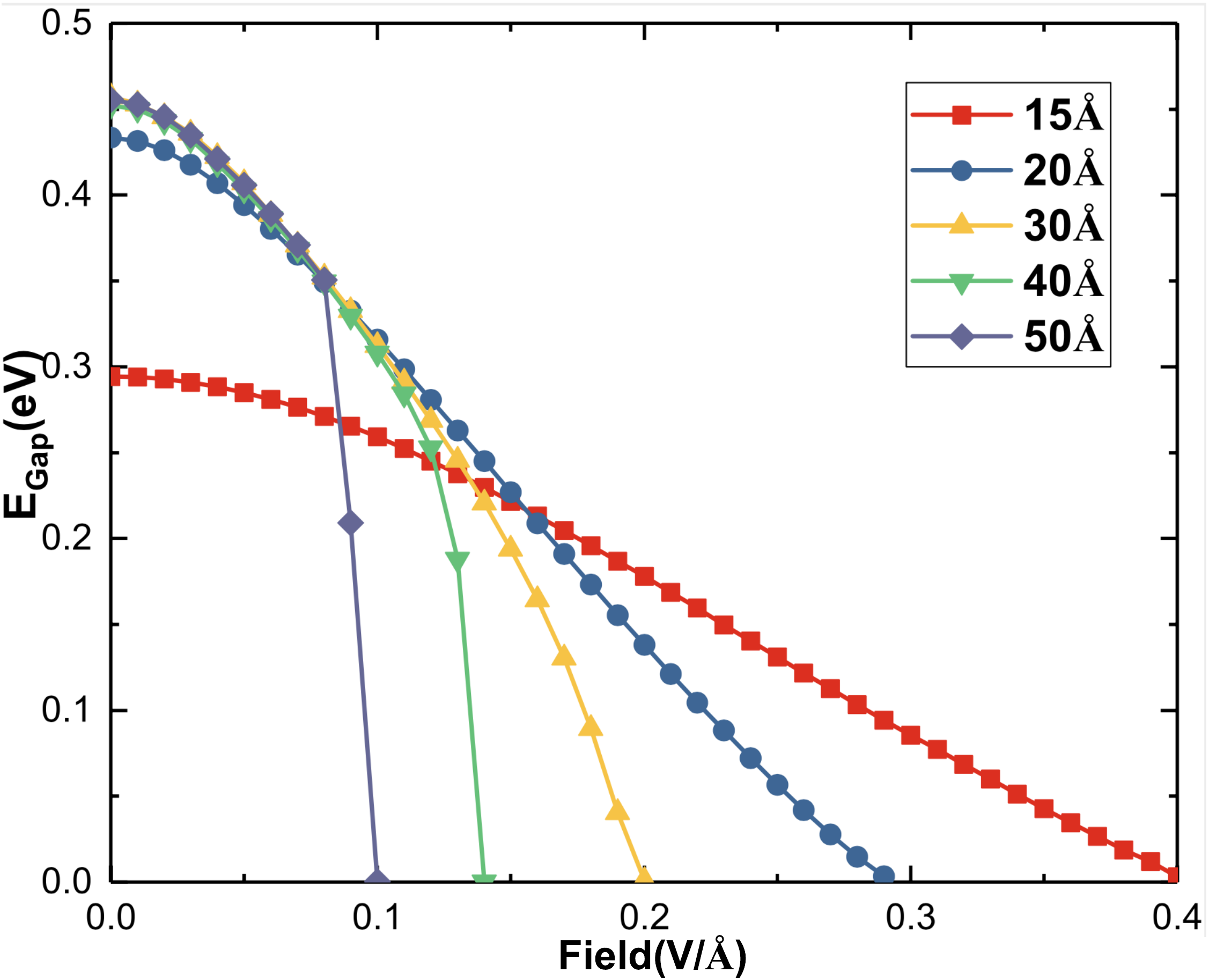}
\caption{J. Zhou \emph{et al.}} \label{gap_mod}
\end{figure}

\begin{figure}[tbp]
\includegraphics[width=1\textwidth]{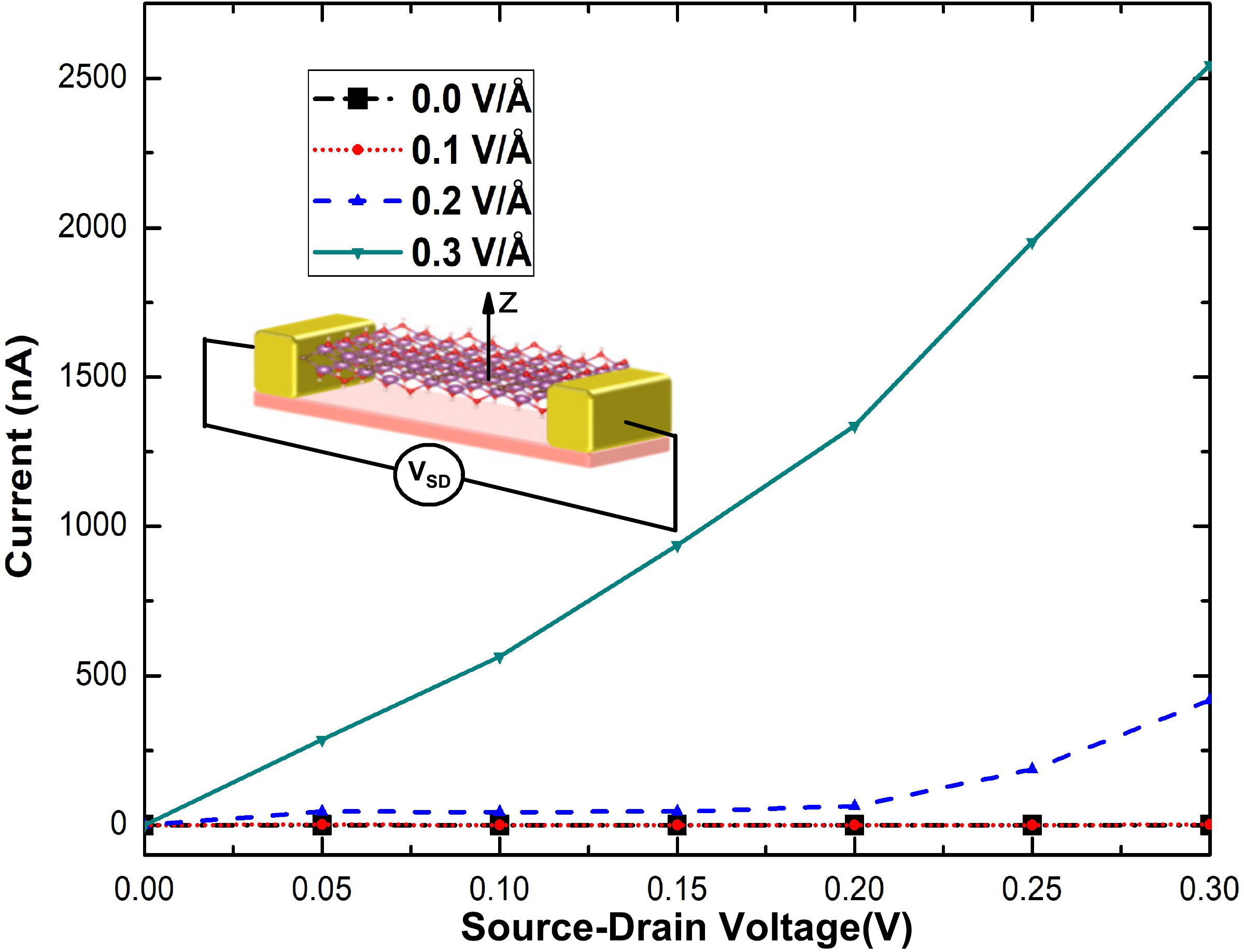}
\caption{J. Zhou \emph{et al.}} \label{IV}
\end{figure}

\begin{description}
\item{FIG. \ref{structure_OH}} Three stable structures of Sc$_2$C(OH)$_2$ monolayers. (a) and (b) are the BB-type structure with different views. (c) and (d) are AB- and AA-type model structures, respectively. The box in (a) indicates the primitive cell.

\item{FIG. \ref{phonon}} (a) cNEB calculation of three stable structures. (b)-(d) Phonon spectra for BB, AB and AA model structures, respectively.

\item{FIG. \ref{fat_bands}} (a)-(c) The fat bands of BB-Sc$_2$C(OH)$_2$, AB-Sc$_2$C(OH)$_2$ and AA-Sc$_2$C(OH)$_2$, respectively. The red curves show the energy bands E$_{\vec{k}}$. The size of the green dots indicates the summation of the weights of the projected wave functions onto the atomic orbitals (\textit{s}+\textit{p}+\textit{d} orbitals). E$_{WF}$ is the work function with respect to the Fermi level.

\item{FIG. \ref{pdos}} The fat bands of valence bands around the $\Gamma$-point. (a) $d$ orbitals of two Sc atoms. (b) $p$ orbitals of Sc atoms. (c) $p$ orbitals of C atom. (d) Projected DOS onto all atoms. The inset shows the projected dos of two lowest NFE states.

\item{FIG. \ref{density}} The averaged decomposed density of six lowest NFE bands at $\Gamma$-point along the $z$-axis. Only results of BB-structure are plotted. ”$+$” and ”$-$” indicate even and odd parity, respectively. The decomposed densities are plotted below. 

\item{FIG. \ref{WaveNFE}} (a)-(b) The wave functions related to image-potential model. (c) The wave functions related to the square well with the same width and depth.

\item{FIG. \ref{Rydberg}} The linear fitting of eigenvalues obtained using the image-potential model.

\item{FIG. \ref{bands}} The energy bands with different vacuum sizes.

\item{FIG. \ref{densNFE}} The charge densities of the NFE states obtained using the image-potential model with periodical neighbors.

\item{FIG. \ref{gap_mod}} The responses of energy gap widths to the electric fields and their dependence on the interlayer distances.

\item{FIG. \ref{IV}} The \textit{IV} curves represent the current with respect to source-drain voltage. Different gate fields (0.0, 0.1, 0.2 and {0.3~V/\AA}, along $z$ axis) are applied. \end{description}

\clearpage\newpage

\clearpage\newpage

\clearpage\newpage

\clearpage\newpage

\clearpage\newpage

\clearpage\newpage
\begin{figure}[tbp]
\includegraphics[width=1\textwidth]{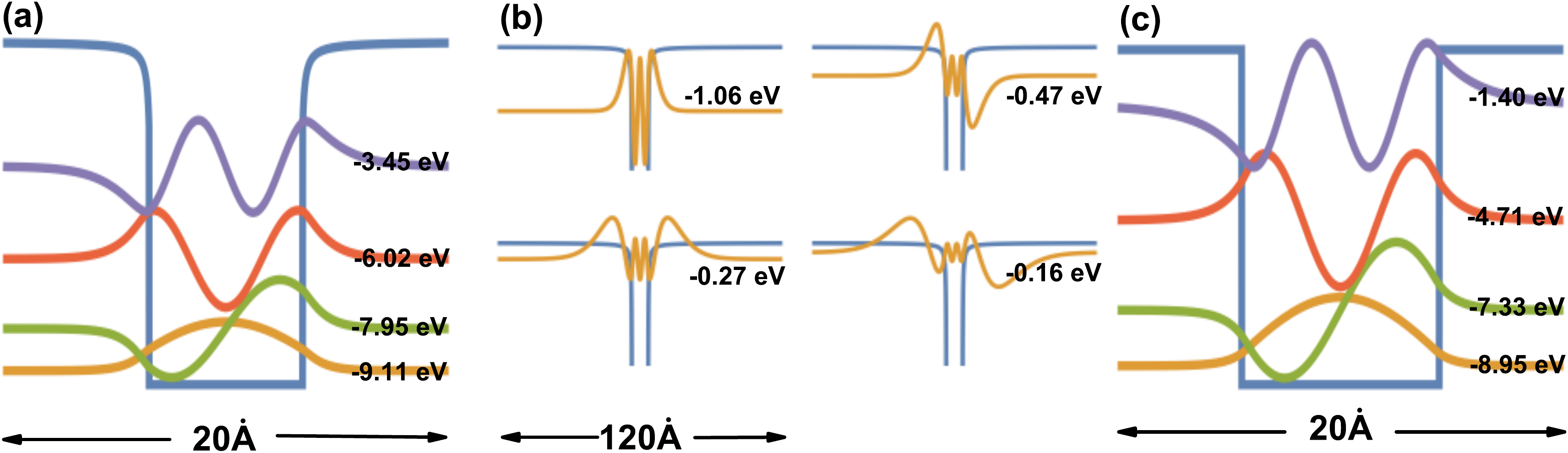}
\caption{J. Zhou \emph{et al.}} \label{WaveNFE}
\end{figure}

\clearpage\newpage

\clearpage\newpage

\clearpage\newpage

\clearpage\newpage

\clearpage\newpage


\end{document}


\title{Modulation of Nearly Free Electron States in Hydroxyl-Functionalized MXenes: A First-Principles Study}

\author{Jiaqi Zhou}
\affiliation{Department of Physics, Shanghai Normal University, Shanghai 200234, China}

\author{Mohammad Khazaei}
\affiliation{Department of Physics, Yokohama National University, Yokohama 240-8501, Japan}

\author{Ahmad Ranjbar}
\affiliation{Dynamics of Condensed Matter and Center for Sustainable Systems Design, Chair of Theoretical Chemistry, University of Paderborn, Warburger Str. 100, D-33098 Paderborn, Germany}

\author{Vei Wang}
\affiliation{Department of Applied Physics, Xi’an University of Technology, Xi’an 710054, China}

\author{Thomas D. K\"{u}hne}
\affiliation{Dynamics of Condensed Matter and Center for Sustainable Systems Design, Chair of Theoretical Chemistry, University of Paderborn, Warburger Str. 100, D-33098 Paderborn, Germany}

\author{Kaoru Ohno}
\affiliation{Department of Physics, Yokohama National University, Yokohama 240-8501, Japan}

\author{Yoshiyuki Kawazoe}
\affiliation{New Industry Creation Hatchery Center, Tohoku University, Sendai, 980-8579, Japan}
 \affiliation{School of Physics, Institute of Science and Center of Excellence in Advanced Functional Materials, Suranaree University of Technology, Nakhon Ratchasima 30000, Thailand}

\author{Yunye Liang}
\email{E-mail: liangyunye@shnu.edu.cn}
\affiliation{Department of Physics, Shanghai Normal University, Shanghai 200234, China}

\date{\today}

\widetext

\maketitle
\section{The solutions of image-potential well}


The Schr\"{o}dinger equation is
\begin{equation}
-\frac{\hbar^2}{2m}\frac{d^2}{dz^2}\psi(z)+V(z)\psi(z)=E\psi(z)
\end{equation} where
\begin{numcases}{V(z)=}
-V_0 & $|z| \leq a$\\
-\frac{e^2}{4(|z|-a+\Delta)} & $|z| > a$
\end{numcases}
and $\Delta=\frac{e^2}{4V_0}$. 
The eigenstates of the Schr\"{o}dinger equation are solved in different regions, respectively. 

If $z<-a$, the above equation becomes
\begin{equation}
\label{less_negative_a}
-\frac{\hbar^2}{2m}\frac{d^2}{dz^2}\psi(z)+\frac{e^2}{4(z+a-\Delta)}\psi(z)=E\psi(z)
\end{equation}
Let $a_0=\frac{\hbar^2}{me^2}$ the Bohr's radius, $E_0=\frac{e^2 }{2a_0}=13.6 $ and $\frac{E}{E_0}=\zeta$. Eq. \ref{less_negative_a} is 
\begin{equation}
\label{psi}
\frac{d^2\phi(\eta)}{d\eta^2}-\frac{1}{2\eta}\phi(\eta)+\zeta\phi(\eta)=0,
\end{equation}
where $\frac{z+a-\Delta}{a_0}=\eta$ and $\psi(z)=\psi(a_0\eta -a+\Delta)=\phi(\eta)$.
Only bound states are considered, therefore $\zeta<0$. Substitute 
$\phi(\eta)=e^{\sqrt{-\zeta}\eta}\eta f(\eta)$ into Eq. \ref{psi}, we get
\begin{equation}
2\eta \frac{d^2 f(\eta)}{d\eta^2}+4(1+\eta \sqrt{-\zeta})\frac{df(\eta)}{d\eta}+(4\sqrt{-\zeta}-1)f(\eta)=0
\end{equation}
If we assume $\eta=-\frac{1}{2\sqrt{-\zeta}}\rho$ and $f(\eta)=f(-\frac{1}{2\sqrt{-\zeta}}\rho)=F(\rho)$, the above equation becomes
\begin{equation}
\rho \frac{d^2F(\rho)}{d\rho^2}+(2-\rho)\frac{dF(\rho)}{d\rho}-(1-\frac{1}{4\sqrt{-\zeta}})F(\rho)=0
\end{equation}
This is confluent hypergeometric equation or Kummer's equation,~\cite{kummer_equation} which has two independent solutions.
\begin{equation}
F(\rho)=A U(1-\frac{1}{4\sqrt{-\zeta}},2,\rho)+B _1F_1(1-\frac{1}{4\sqrt{-\zeta}},2,\rho)
\end{equation}
where $A$ and $B$ are constants to be determined, $_1F_1(1-\frac{1}{4\sqrt{-\zeta}},2,\rho)$ is the Kummer's function~\cite{kummer_equation} and this solution is nonsense because $e^{\sqrt{-\zeta}\eta}\eta_1F_1(1-\frac{1}{4\sqrt{-\zeta}},2,-2\sqrt{-\zeta}\eta)$ is infinitive if $\eta \to -\infty$. Thus the only possible solution is $U(1-\frac{1}{4\sqrt{-\zeta}},2,\rho)$, which is Tricomi's function.~\cite{kummer_equation} The finial solution  of Eq. \ref{less_negative_a} is 
\begin{equation}
\psi_1(z)=A e^{\frac{\sqrt{-\zeta}(z+a-\Delta)}{a_0}}\frac{z+a-\Delta}{a_0} U\left (1-\frac{1}{4\sqrt{-\zeta}},2,\frac{-2\sqrt{-\zeta}(z+a-\Delta)}{a_0} \right).
\end{equation} where $A$ is a constant to be determined.

In the middle area ($|z| \leq a$), the potential is $-V_0$ and the Schr\"{o}dinger equation is
\begin{equation}
-\frac{\hbar^2}{2m}\frac{d^2}{dz^2}\psi(z)-V_0(z)\psi(z)=E\psi(z)
\end{equation}
Using the Bohr's radius and Rydberg unit of energy in the above equation and let $\eta=\frac{z}{a_0}$, we get
\begin{equation}
\frac{d^2\phi(\eta)}{d\eta^2}+k^2\phi(\eta)=0
\end{equation} where $k^2=\frac{E+V_0}{E_0}$ and the solution is
\begin{equation}
\psi_2(z)=B \sin(k\frac{z}{a_0})+C \cos(k\frac{z}{a_0})
\end{equation} where $B$ and $C$ are constants to be determined.

If $z>a$, the solving process is very similar to the previous one and the Schr\"{o}dinger equation reads
\begin{equation}
\label{great_positive_a}
-\frac{\hbar^2}{2m}\frac{d^2}{dz^2}\psi(z)-\frac{e^2 }{4(z-a+\Delta)}\psi(z)=E\psi(z)
\end{equation}
By using Bohr's radius and Rydberg unit of energy, we find
\begin{equation}
\label{psi2}
\frac{d^2\phi(\eta)}{d\eta^2}+\frac{1}{2\eta}\phi(\eta)+\zeta\phi(\eta)=0
\end{equation}
where $\frac{z-a+\Delta}{a_0}=\eta$ and $\psi(z)=\psi(a_0\eta +a-\Delta)=\phi(\eta)$.
By comparison, we find we can get Eq. \ref{psi} from Eq. \ref{psi2} by the substitution $\eta \to -\eta$. Therefore, the corresponding solution is 
\begin{equation}
\psi_3(z)=D e^{\frac{-\sqrt{-\zeta}(z-a+\Delta)}{a_0}}\frac{z-a+\Delta}{a_0} U\left (1-\frac{1}{4\sqrt{-\zeta}},2,\frac{2\sqrt{-\zeta}(z-a+\Delta)}{a_0} \right).
\end{equation} where $D$ is an undetermined constant.

Given the fact that the potential well is symmetric, the wave functions have even or odd parities, respectively. As a result, only two undetermined constants are independent. For example, if the wave function is even, $B=0$ and $D=-A$. The constants $A$ and $C$ can be determined when the conditions on continuity of the wave function and its derivatives are applied, i.e.
\begin{gather}
\psi_1(-a)=\psi_2(-a) \\
\frac{d\psi_1(-a)}{dz}=\frac{d\psi_2(-a)}{dz} \
\end{gather}
With the differentiation equality $\frac{dU(a,b,z)}{dz}=-aU(a+1,b+1,z)$, the above equations are equivalent to the following matrix multiplication,
\begin{gather}
\label{matrixM}
M^e \times V=
\begin{pmatrix}
M_{11}  &  -\cos(k\frac{a}{a_0}) \\
M_{21} &  -\frac{k}{a_0}\sin(k\frac{a}{a_0})\
\end{pmatrix}
\times
\begin{pmatrix}
A \\ C \
\end{pmatrix}
=0
\end{gather}
where
\begin{align}
M_{11}&=-\frac{\Delta}{a_0}e^{-\frac{\sqrt{-\zeta}\Delta}{a_0}} U\left (1-\frac{1}{4\sqrt{-\zeta}},2,\frac{2\sqrt{-\zeta}\Delta}{a_0} \right) \\
M_{21}&=\frac{e^{-\frac{\Delta  \sqrt{-\zeta }}{{a_0}}} \left(2 \left({a_0}-\Delta 
   \sqrt{-\zeta }\right) U\left(1-\frac{1}{4 \sqrt{-\zeta }},2,\frac{2 \Delta 
   \sqrt{-\zeta }}{{a_0}}\right)+\Delta  \left(1-4 \sqrt{-\zeta }\right)
   U\left(2-\frac{1}{4 \sqrt{-\zeta }},3,\frac{2 \Delta  \sqrt{-\zeta
   }}{{a_0}}\right)\right)}{2 {a_0}^2}
\end{align}
The determinant of matrix $M^e$ (namely Wronskian) is energy-dependent. The constants are nonzero if and only if the determinant of matrix $M^e$  is zero. The corresponding energies indicate the binding energies. 

 If the wave function has the odd parity, $C=0$ and $D=A$, and Eq.\ref{matrixM} becomes
 \begin{gather}
\label{matrixM}
M^o \times V=
\begin{pmatrix}
M_{11}  &  \sin(k\frac{a}{a_0}) \\
M_{21} &  -\frac{k}{a_0}\cos(k\frac{a}{a_0})\
\end{pmatrix}
\times
\begin{pmatrix}
A \\ B \
\end{pmatrix}
=0
\end{gather}

\newpage

\begin{flushleft}
{\large \textbf{Captions} }
\end{flushleft}

\begin{figure}[tbp]
\includegraphics[width=1\textwidth]{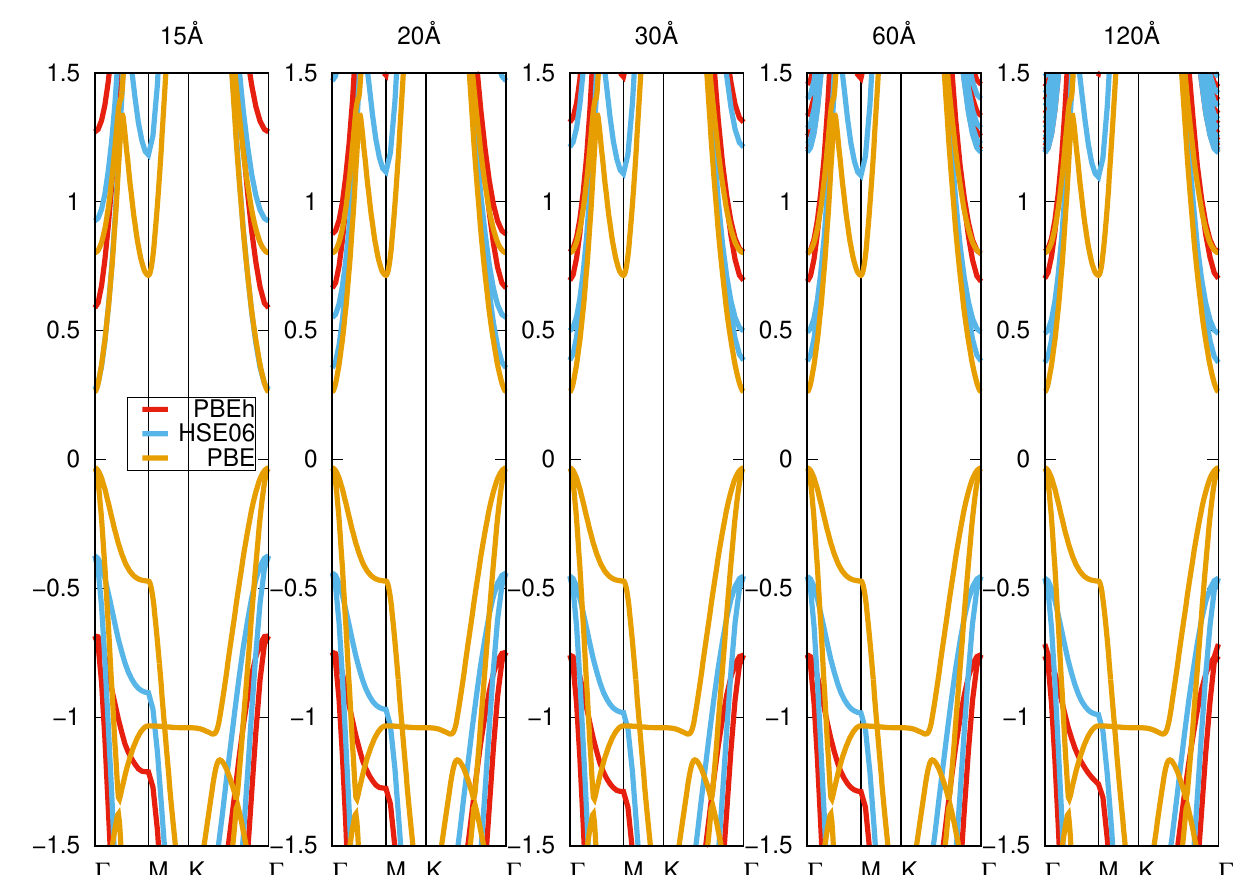}
\caption{J. Zhou \emph{et al.}} \label{pbe0}
\end{figure}

\begin{figure}[tbp]
\includegraphics[width=1\textwidth]{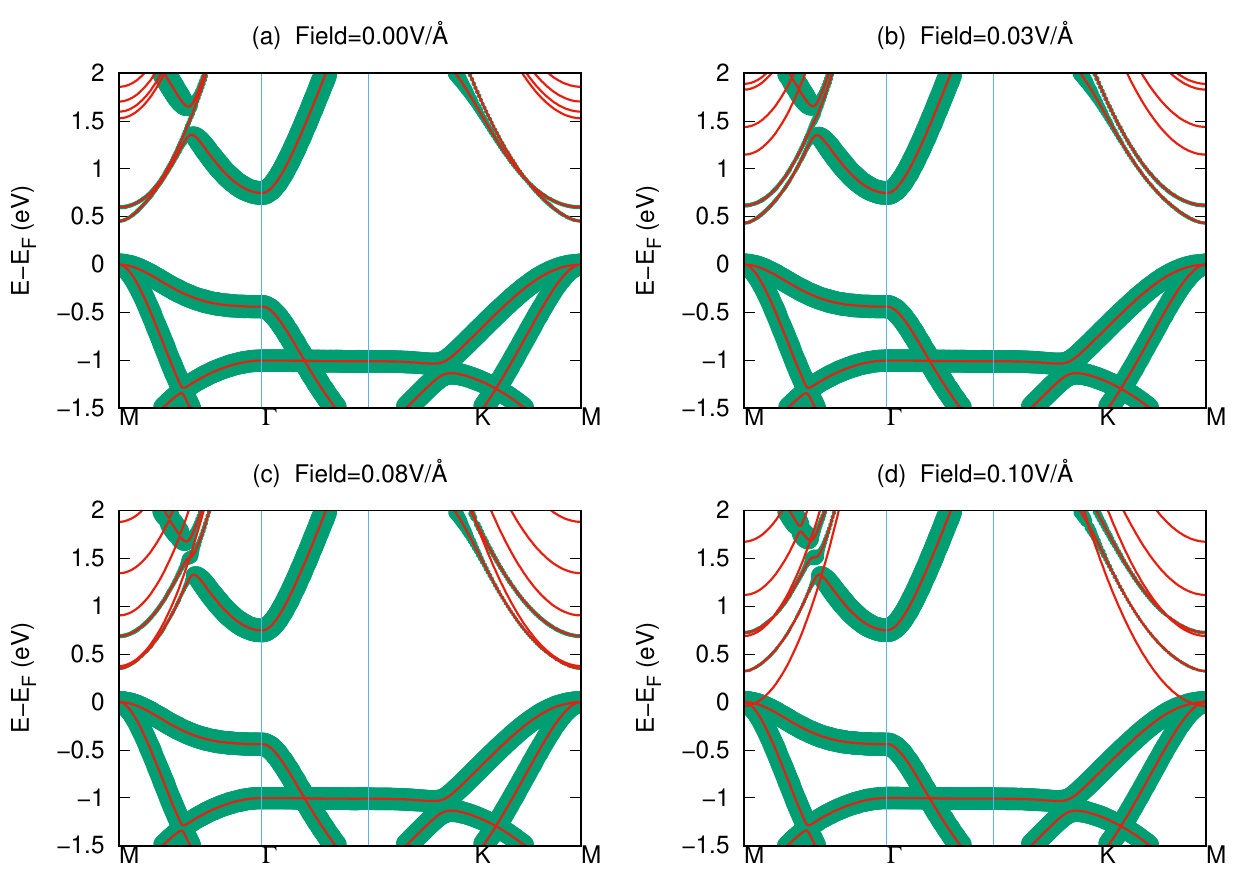}
\caption{J. Zhou \emph{et al.}} \label{50A}
\end{figure}

\begin{figure}[tbp]
\includegraphics[width=1\textwidth]{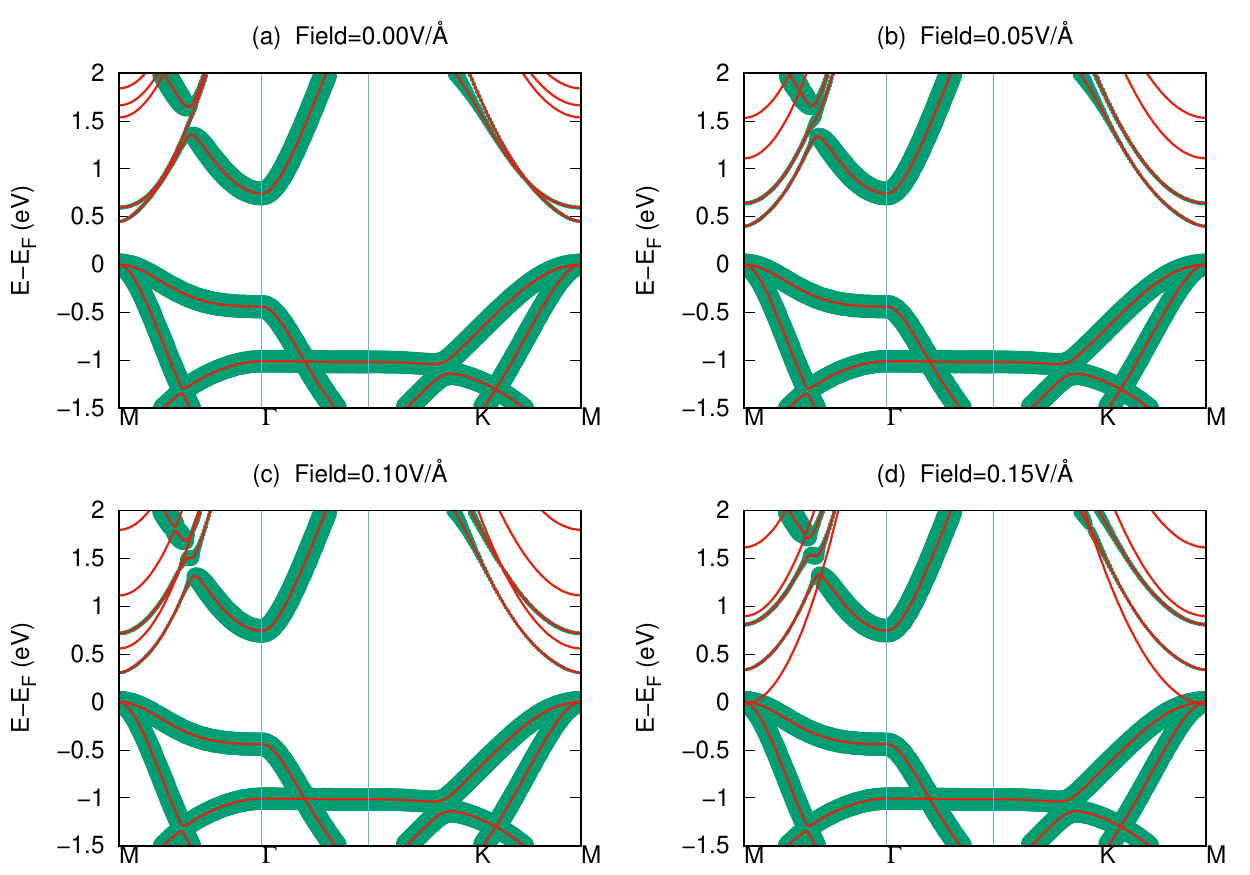}
\caption{J. Zhou \emph{et al.}} \label{40A}
\end{figure}

\begin{figure}[tbp]
\includegraphics[width=1\textwidth]{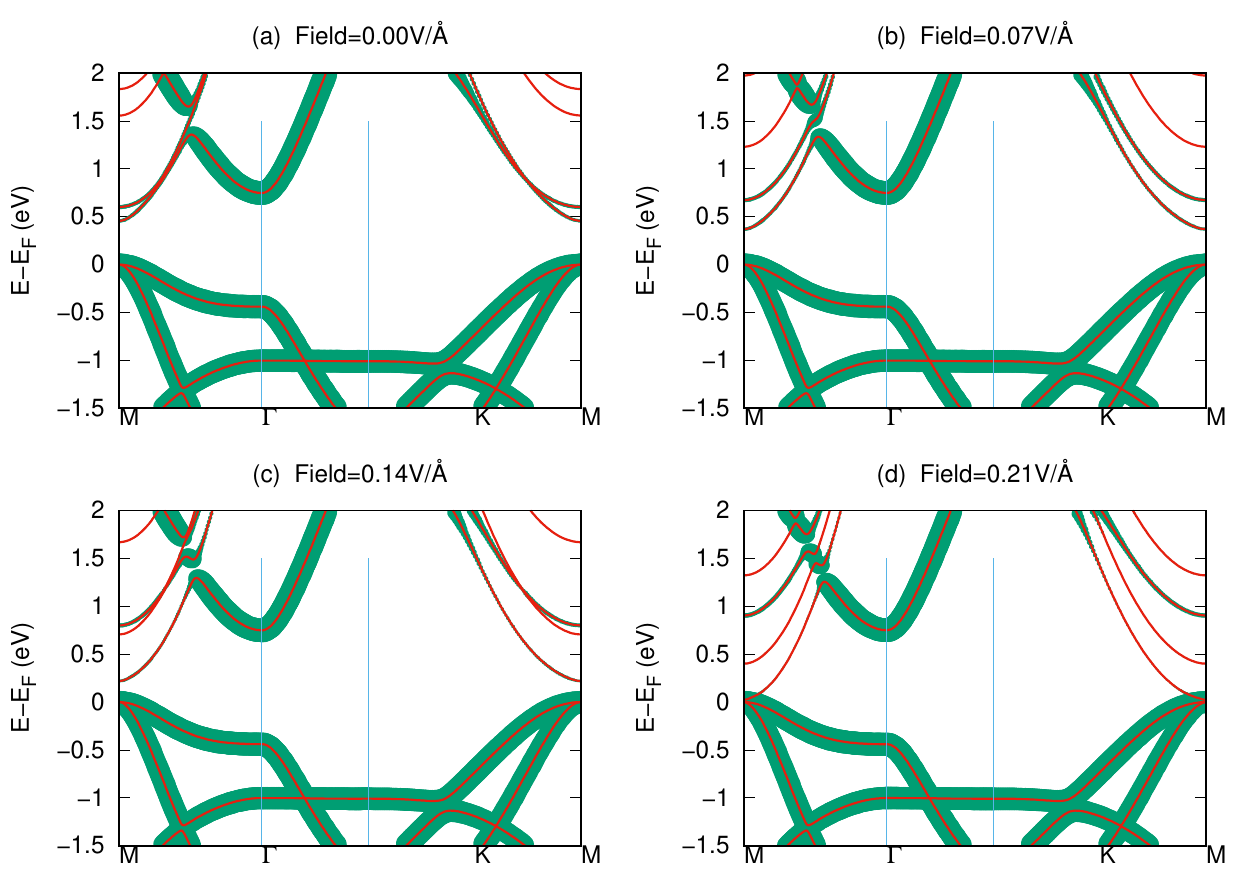}
\caption{J. Zhou \emph{et al.}} \label{30A}
\end{figure}

\begin{figure}[tbp]
\includegraphics[width=1\textwidth]{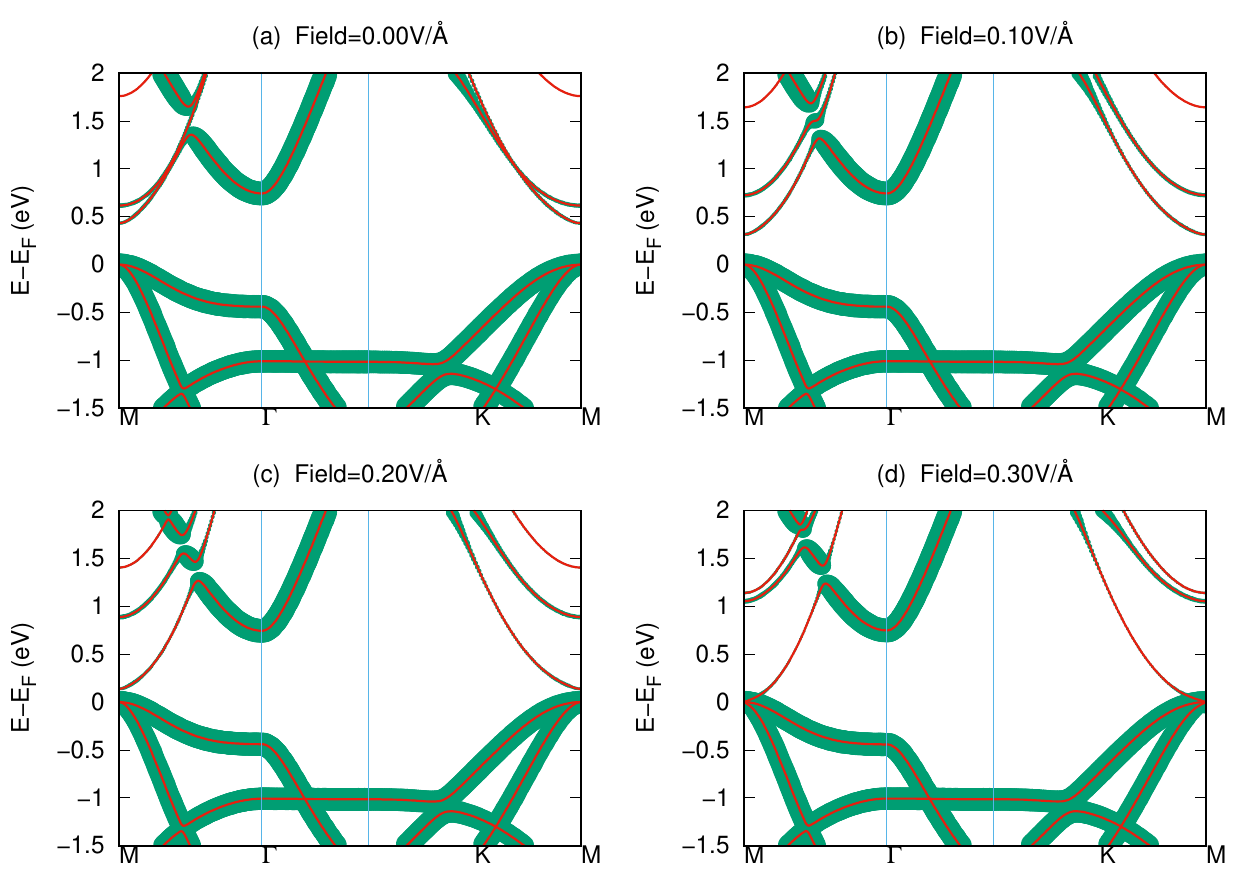}
\caption{J. Zhou \emph{et al.}} \label{20A}
\end{figure}

\begin{figure}[tbp]
\includegraphics[width=1\textwidth]{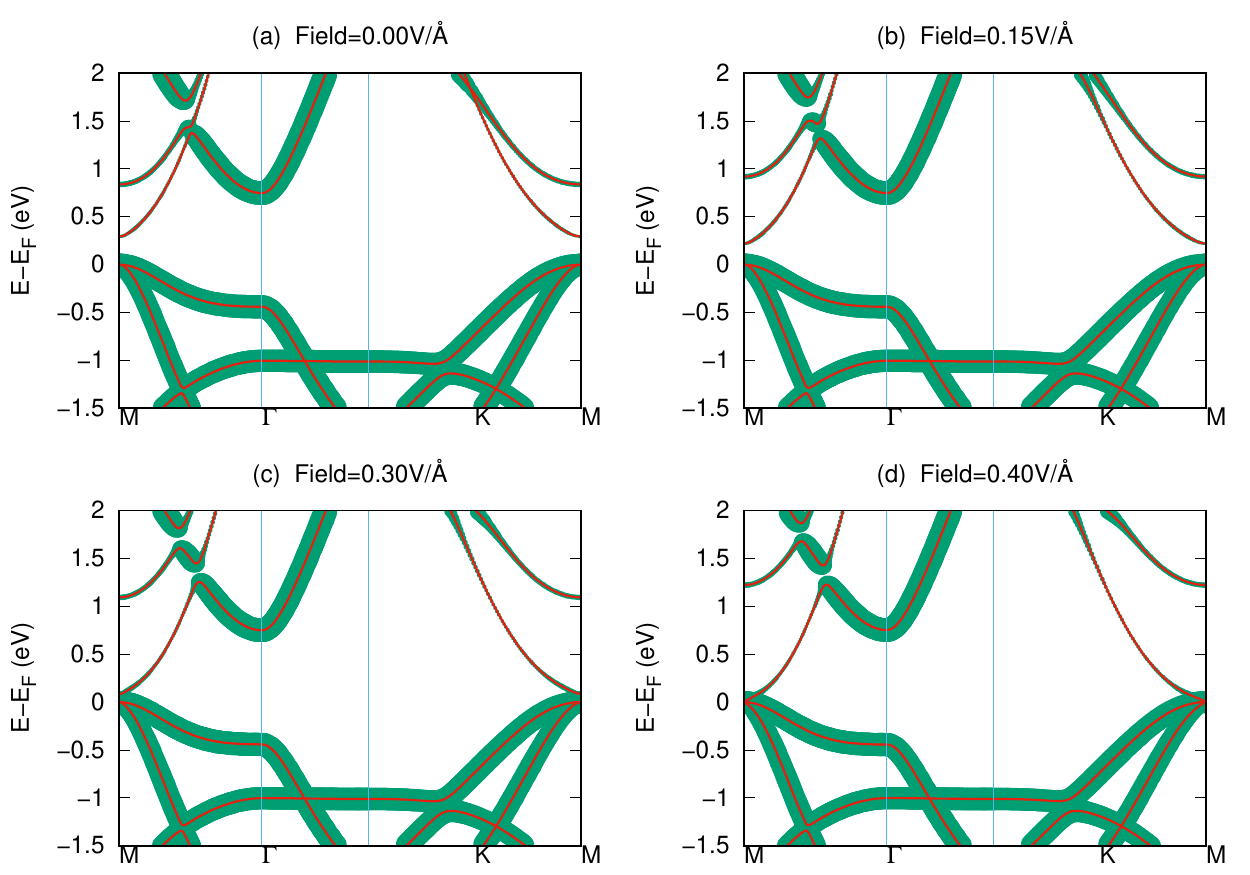}
\caption{J. Zhou \emph{et al.}} \label{15A}
\end{figure}

\begin{figure}[tbp]
\includegraphics[width=1\textwidth]{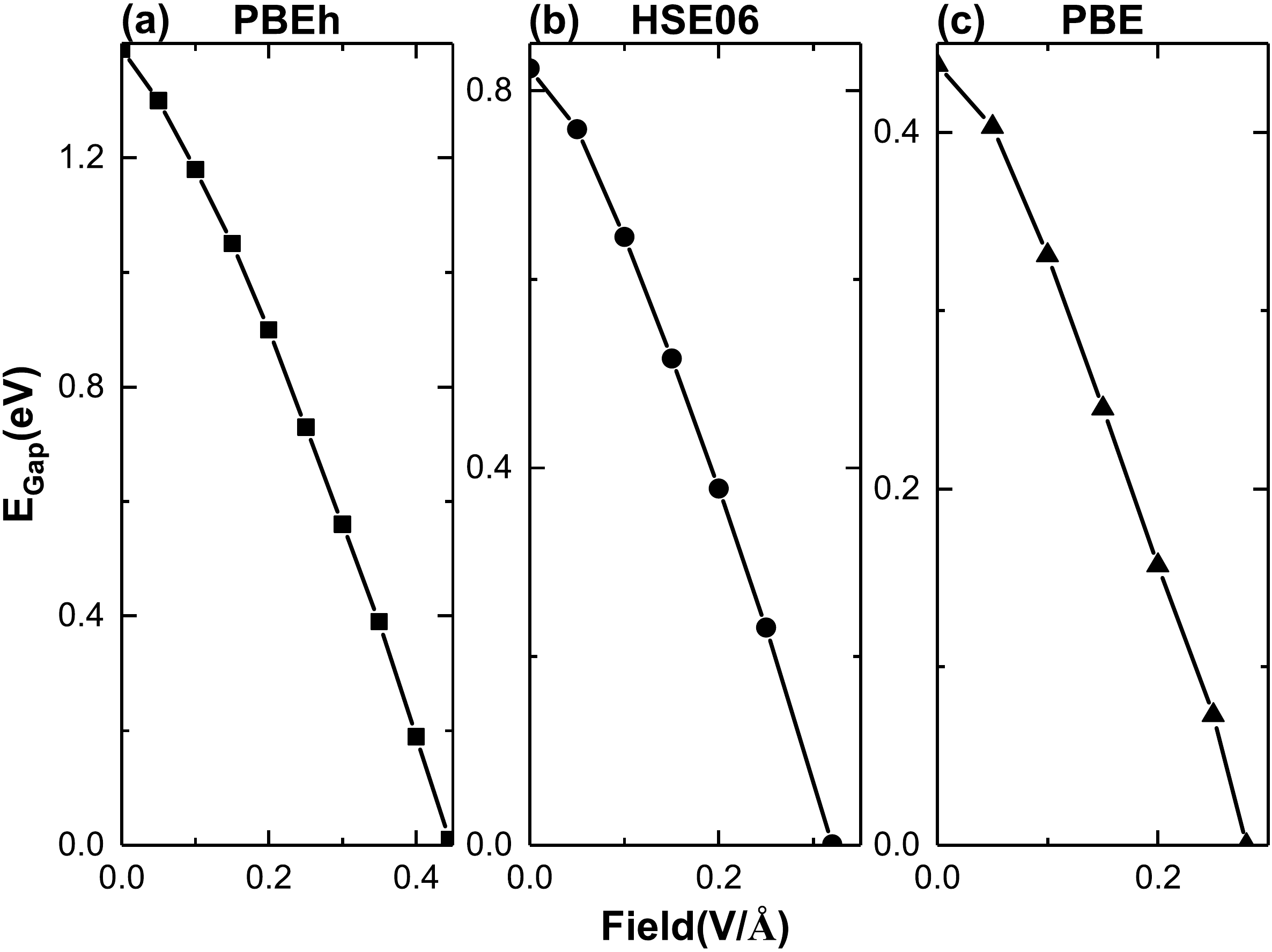}
\caption{J. Zhou \emph{et al.}} \label{bands_pbeh_hse06}
\end{figure}

\begin{figure}[tbp]
\includegraphics[width=1\textwidth]{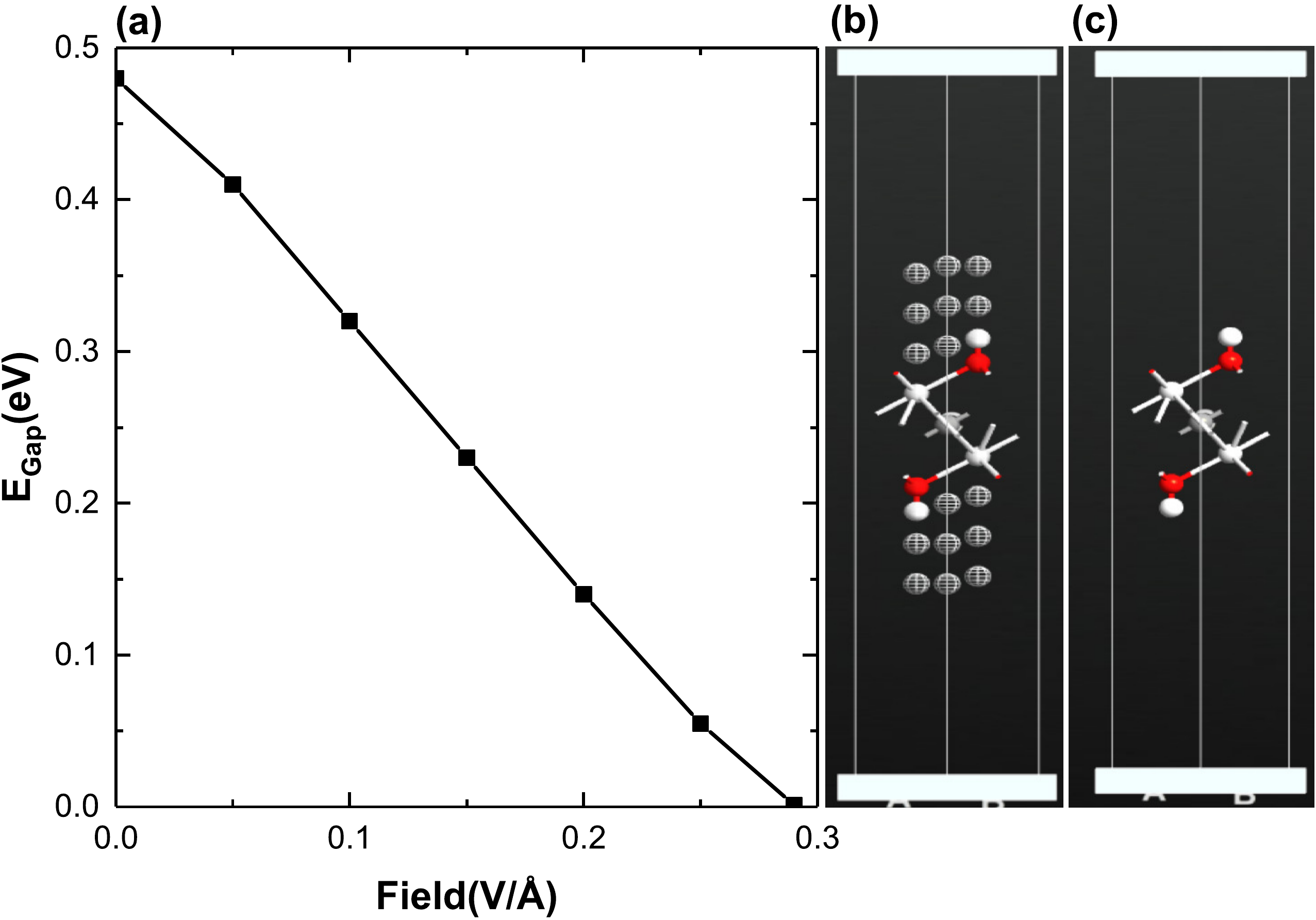}
\caption{J. Zhou \emph{et al.}} \label{ghost_atom}
\end{figure}

\begin{description}
\item{FIG. \ref{pbe0}} The energy bands with respect to different interlayer distances were calculated by using PBEh (red), HSE06 (blue) and PBE (yellow).
\item{FIG. \ref{50A}} The response of energy bands to the electric fields. The distance between the neighboring layers is 50~\AA. When the field is about 0.08eV/\AA, the exchange of NFE states happens. 
\item{FIG. \ref{40A}} The response of energy bands to the electric fields. The distance between the neighboring layers is 40~\AA.
\item{FIG. \ref{30A}} The response of energy bands to the electric fields. The distance between the neighboring layers is 30~\AA.
\item{FIG. \ref{20A}} The response of energy bands to the electric fields. The distance between the neighboring layers is 20~\AA.
\item{FIG. \ref{15A}} The response of energy bands to the electric fields. The distance between the neighboring layers is 15~\AA.
\item{FiIG. \ref{bands_pbeh_hse06}} The energy band gap widths responses to the electric fields, calculated by PBEh (a), HSE06 (b) and PBE (c). The interlayer separation is 20~{\AA}, and all results are calculated by VASP.
\item{FIG. \ref{ghost_atom}} (a) The calculated energy gap widths responses to the electric fields by ATK, and the distance between the neighboring layers is 20~\AA. The structures used in calculations with (b) and without (c) ghost atoms. The grey atoms represent the ghost atoms.
\end{description}

\newpage
\clearpage\newpage
\begin{table}[tb]
\caption{The relative energies (in the unit of eV) of three stable structures (BB, AB and AA) calculated from PBEh and HSE06, respectively. BB structure is the reference structure and all energies are relative to it.}\label{stablestructure}
\begin{tabular*}{0.5\textwidth}{@{\extracolsep{\fill}}c|cccccc}
 	\hline
 	\hline
sites of OH & BB & BA & AA & TB & TA & TT \\
         \hline
PBEh & 0.000 & 0.422 & 0.883  & 1.564 & 2.081 & 2.818 \\
 	\hline
HSE06 & 0.000 & 0.347 & 0.766 & 1.557 & 2.072 & 2.806\\
 	\hline
 	\hline
\end{tabular*}
\end{table}

\begin{table}[tb]
\caption{The band gap width (in the unit of eV) evolution with respect to the vacuum size (in the unit of \AA).}\label{band_gap_vacuum}
\begin{tabular*}{0.5\textwidth}{@{\extracolsep{\fill}}c|cccccccc}
 	\hline
 	\hline
vacuum size({\AA}) & 120 & 60 & 30 & 20 & 18 & 15 \\
         \hline
gap width(eV) & 0.46 & 0.46 & 0.45  & 0.43 & 0.40 & 0.30 \\
 	\hline
 	\hline
\end{tabular*}
\end{table}

\newpage
\clearpage\newpage

\clearpage\newpage

\clearpage\newpage

\clearpage\newpage

\clearpage\newpage

\clearpage\newpage

\clearpage\newpage

\clearpage\newpage
